\documentclass{SCIS2014}

\usepackage{mathrsfs}

\usepackage{multirow} 

\begin{document}

\ArticleType{RESEARCH PAPER}
\Year{2015}
\Month{January}
\Vol{58}
\No{1}
\DOI{0.1007/s11432-014-5237-y}
\ArtNo{011101}
\ReceiveDate{}
\AcceptDate{}
\OnlineDate{}

\title{Link Prediction in Social Networks: \\ the State-of-the-Art}{Link Prediction in Social Networks: the State-of-the-Art}

\author[1,3]{WANG Peng}{pwang@seu.edu.cn}
\author[1,2,3]{XU BaoWen}{bwxu@nju.edu.cn}
\author[1]{WU YuRong}{}
\author[1]{ZHOU XiaoYu}{}

\AuthorMark{Wang P, Xu BW}

\AuthorCitation{WANG Peng, XU BaoWen, WU YuRong, ZHOU XiaoYu}

\address[1]{School of Computer Science and Engineering, Southeast University, Nanjing {\rm 210096}, China}
\address[2]{State Key Laboratory for Novel Software Technology, Nanjing University, Nanjing {\rm 210023}, China}
\address[3]{State Key Laboratory of Software Engineering, Wuhan University, Wuhan {\rm 430072}, China}

\maketitle

\abstract{
In social networks, link prediction predicts missing links in current networks and new or dissolution links in future networks, is important for mining and analyzing the evolution of social networks. 
In the past decade, many works have been done about the link prediction in social networks. 
The goal of this paper is to comprehensively review, analyze and discuss the state-of-the-art of the link prediction in social networks.
A systematical category for link prediction techniques and problems is presented.
Then link prediction techniques and problems are analyzed and discussed.
Typical applications of link prediction are also addressed.
Achievements and roadmaps of some active research groups are introduced. 
Finally, some future challenges of the link prediction in social networks are discussed.
}



\section{Introduction}

A social network is a social structure made up of a set of social actors and a set of the ties between these actors. 
A social network can be visualized as a graph, where nodes represent actors/participants (individuals, organizations, et al.) and edges (i.e. links) correspond to ties/interactions/relationships between actors.
With the rapid development of internet, communication and cooperation between people have become more convenient.
In recent years, online social networks such as Facebook, Twitter and Weibo, have become an important part of our daily life and provide us platforms to exchange information with each other.
Since the huge amounts of data on social networks has some obvious characteristics such as high quality, big data, semi-structure and direct reflection of real human society, many researchers from different areas or disciplines pay more and more attention to social networks.
However, mining and analyzing social network data is a non-trivial task, which will face two challenges: incompletion and dynamic.   
First, almost all obtained social network data is incomplete since only part of social information can be collected from social network platforms.
Second, social networks are highly dynamic, that might lead the nodes and edges to appear or disappear in the future. 
Therefore, predicting the missing or unobserved links in current social networks and newly added or deleted links in future social networks is very important, not only for understanding the evolution of social networks, but also for completing current social networks. 
This problem is commonly known as the \textit{Link Prediction}. 
Being one of the link mining and analyzing tasks \cite{GD05}, link prediction has many important applications.
First, it can be applied to recommender systems in information retrieval and e-commerce, 
which can help people to find new friends \cite{ABS12} and potential collaborators \cite{MKK12, WST13}, 
provide interesting items in online shopping \cite{ACF11}, 
 recommend patent partners in enterprise social networks \cite{WST13} and cross-domain partners \cite{TWS12},  find experts or co-authors in academic social networks \cite{PI07, WI08}, 
and predict cell phone contacts in large scale communication network \cite{RLH11}.
Second, it also can be used to infer the complete networks based on partial observed networks \cite{MP08, KL11}, understand the evolution of networks better \cite{BJN02, JMB11, BBB10, RK10}, and predict hyper-links in heterogeneous social networks \cite{ZHH02}. 
Finally, the link prediction techniques can also be applied in bioinformatics and biology, for example, in health care and gene expression networks \cite{AGJ12}, predicting specialists who are more likely to receive future referrals, and finding protein-protein interactions. Even in other domains such as security related domain, it can be used to identify abnormal communications \cite{HJ09}.

\begin{figure}[!t]
\centering
\begin{minipage}[c]{0.32\textwidth}
\centering
\includegraphics[width=0.99\textwidth]{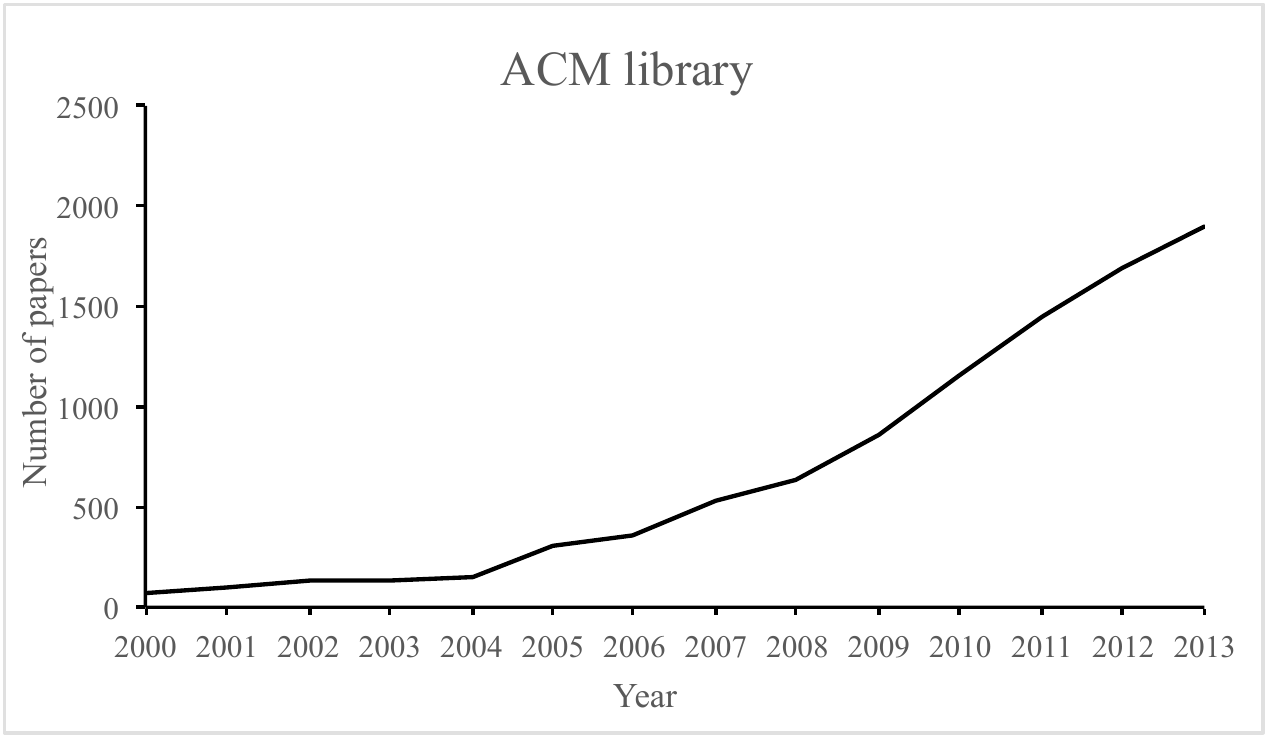}
\end{minipage}
\begin{minipage}[c]{0.32\textwidth}
\centering
\includegraphics[width=0.99\textwidth]{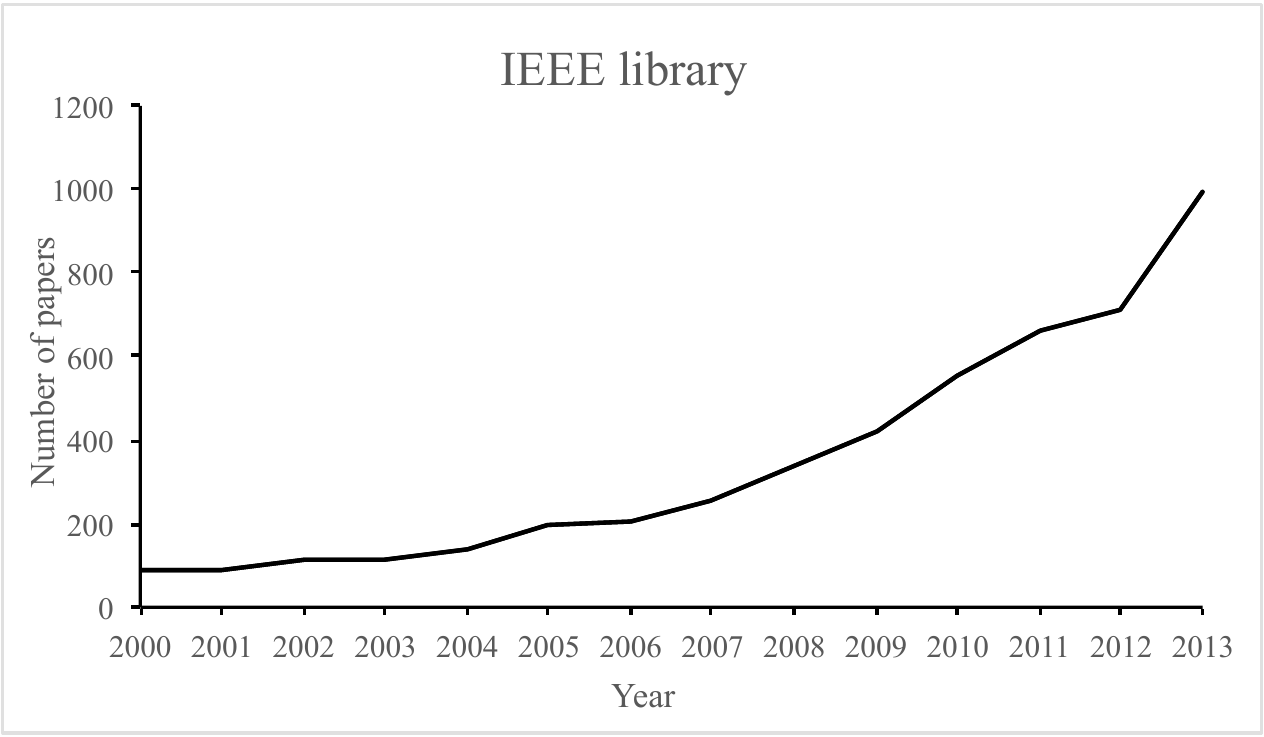}
\end{minipage}
\begin{minipage}[c]{0.32\textwidth}
\centering
\includegraphics[width=0.99\textwidth]{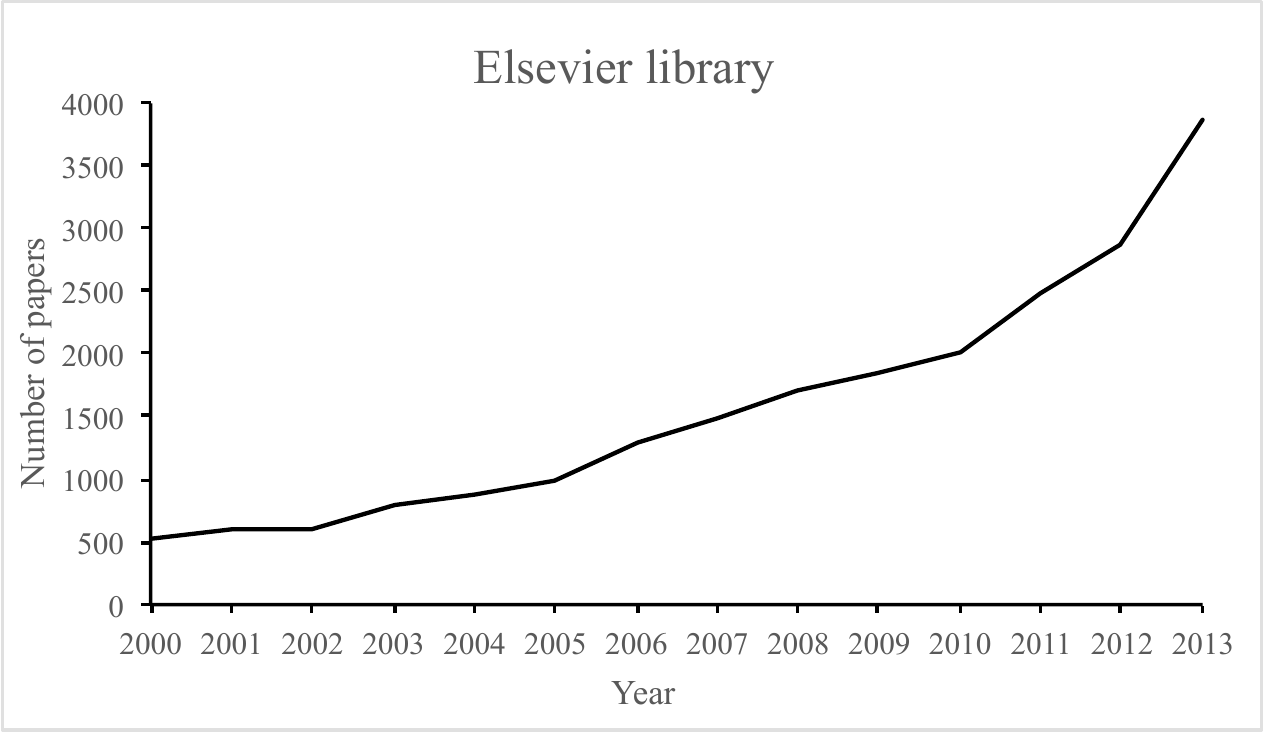}
\end{minipage}
\caption{Published papers related to link prediction problem on ACM, IEEE, and Elsevier libraries.}
\label{fig1}
\end{figure}

\begin{figure}[!t]
\begin{minipage}[c]{0.48\textwidth}
\centering
\includegraphics[width=\textwidth]{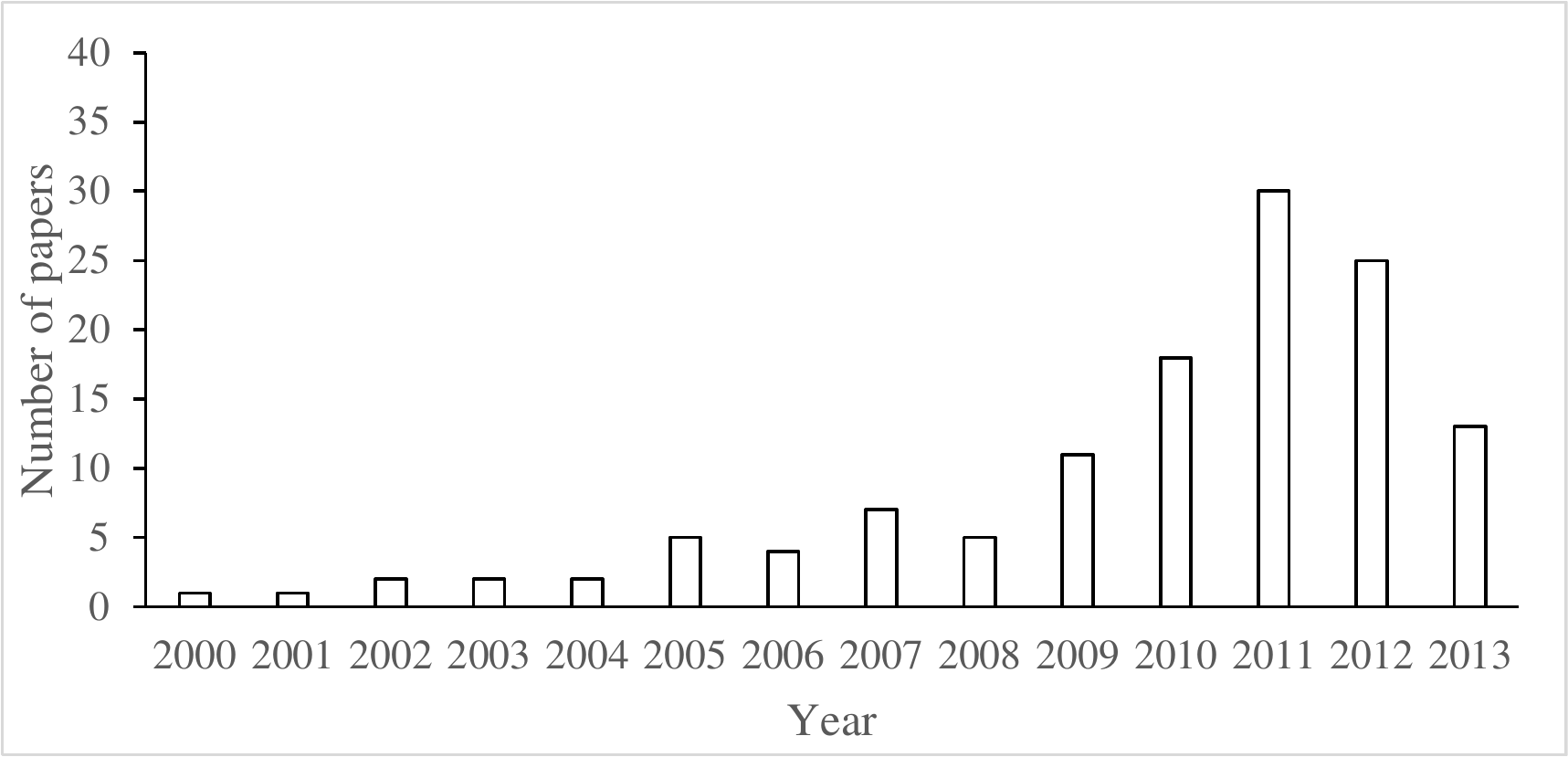}
\caption{Statistics of selected papers by publication date.}
\label{fig2}
\end{minipage}
\begin{minipage}[c]{0.49\textwidth}
\centering
\includegraphics[width=\textwidth]{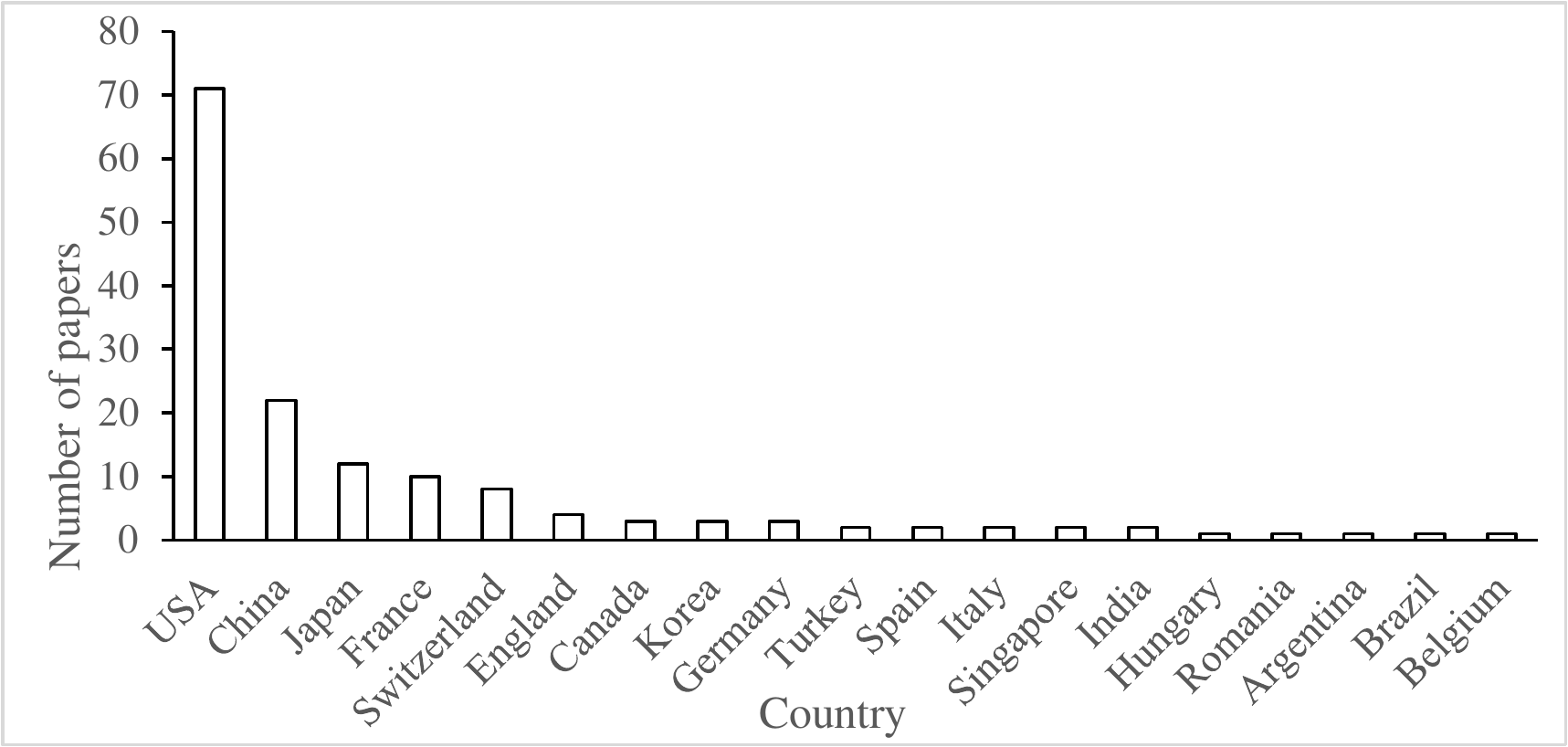}
\caption{Statistics of selected papers by countries.}
\label{fig3}
\end{minipage}
\end{figure}

\begin{table}[!t]
\centering
\caption{Top-10 institutions ranking by number of published papers}
\footnotesize

\begin{tabular*}{0.8\textwidth}{cc}
\toprule
  Institutions & Number of published papers\\
  \hline
Stanford University	& 11\\
Cornell University	& 10\\
University of Fribourg	& 8\\
University of Notre Dame	& 8\\
University of Electronic Science and Technology of China	& 7\\
Tsinghua University	& 7\\
Université Pierre et Marie CURIE	& 5\\
National Institute of Informatics	& 5\\
IBM T. J. Watson Research Center	& 5\\
Yahoo! Research	 & 5\\

\bottomrule
\end{tabular*}
\end{table}

In the past decade, many efforts have been made by psychologists, computer scientists, physicists and economists to solve the link prediction problem in social networks. 
Figure 1 shows the number of published papers with search keywords ``link prediction social network'' on three important computer science libraries: ACM, IEEE, and Elsevier respectively from 2000 to 2013.
It can be seen that more and more works pay attention to the link prediction in social networks, especially in the past five years, there are thousands papers related to this problem every year. 
Therefore, the research trend on this topic is growing.
This paper carefully selects about 130 papers from 2000 to 2013, and most of which are published in the prominent journals (Nature, Physical Review, Physica A, TKDD, TWEB, DKE, JSS, JMLR, Social Networks, Computer Networks, and so on) and conferences (WSDM, SDM, KDD, JCDL, ICDM, CIKM, ICDM, WWW, ISWC, IJCAI, ICML, SIGCOMM, NIPS, ASONAM, et al.). Figure 2 shows the publication date of selected papers, which reveals that most papers are published in the past five years.
Figure 3 shows that these selected papers are mainly finished by 19 countries, and the researchers from USA, China, Japan, France, Switzerland are more active in this problem.
Table 1 lists the top-10 institutions ranking by the number of published papers on link prediction, and we can see that not only academic institutions such as Stanford university and Cornell university, but also industrial companies such as IBM and Yahoo, are active in this problem.
These facts also reflect that the link prediction in social network problem have attracted more and more attention from academic and industrial researchers all over the world in recent years.
Another interesting phenomenon is that link prediction problem also attracts the attentions from different disciplines. 
Figure 4 illustrates the disciplines of all authors in the selected papers and it is found that 98\% authors are from four disciplines: computer science (87\%), physics (7\%), economics (3\%), and management (1\%). 
In particular, according to Figure 5, only 15 (about 10\%) papers are finished by cross-discipline authors: authors of 6 papers are from computer science and physics, authors of 3 papers are from computer science and economics, and 5 papers have co-authors from computer science and other disciplines. Therefore, researchers from different disciplines are still independent and lack cooperation on this problem.  

\begin{figure}[!t]
\centering
\begin{minipage}[c]{0.49\textwidth}
\centering
\includegraphics[width=0.9\textwidth]{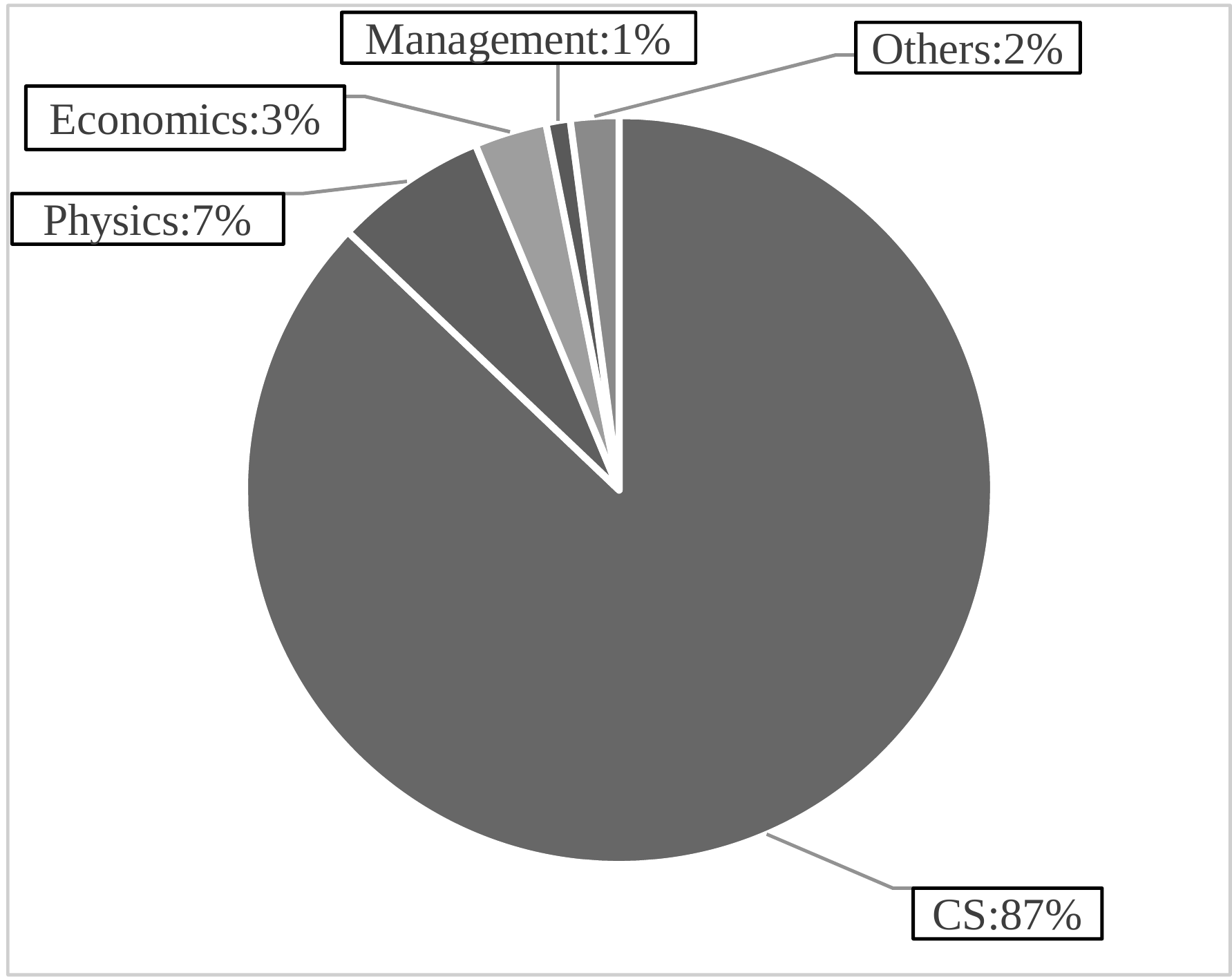}
\end{minipage}
\begin{minipage}[c]{0.49\textwidth}
\centering
\includegraphics[width=0.9\textwidth]{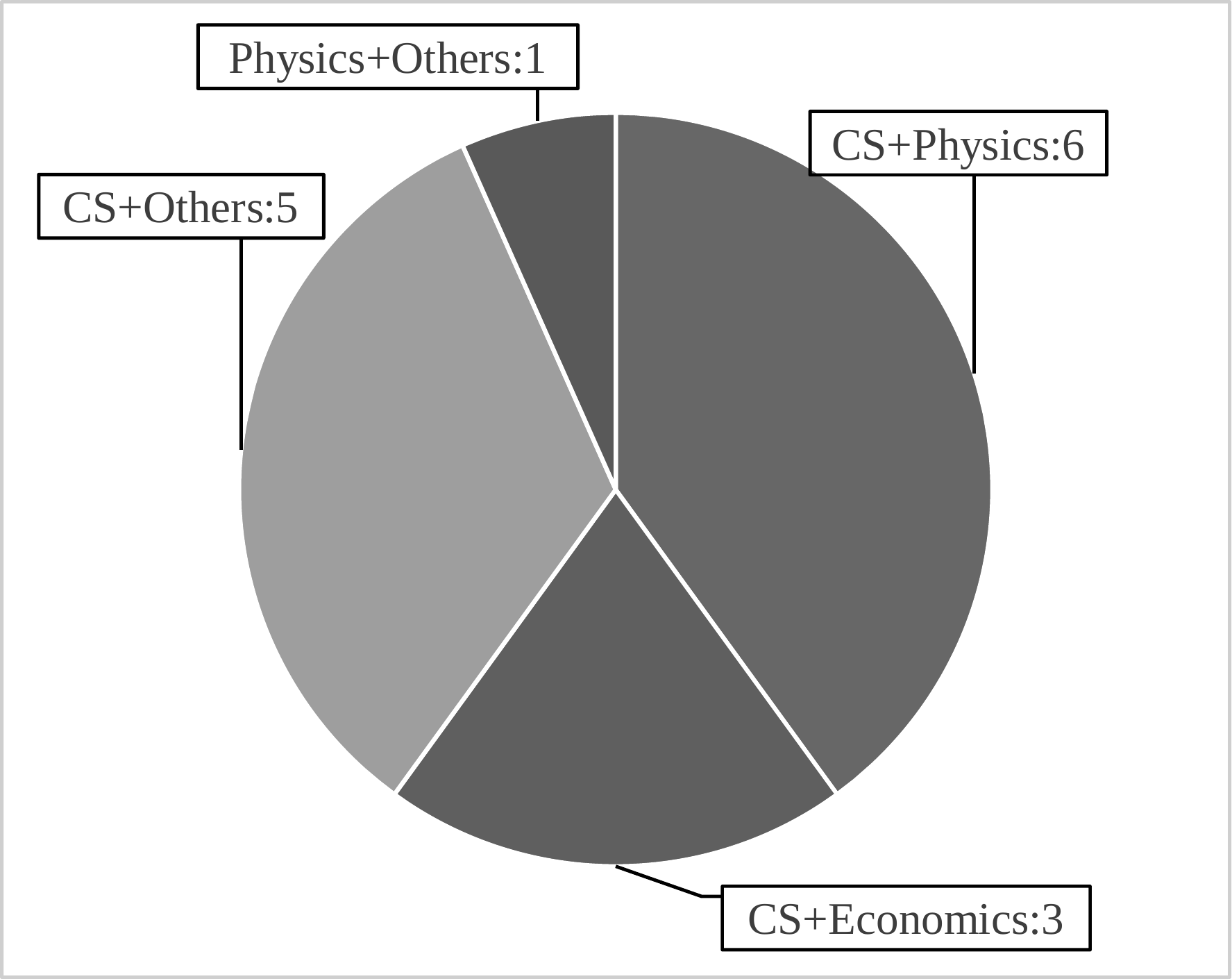}
\end{minipage}
\begin{minipage}[t]{0.49\textwidth}
\centering
\caption{ Statistics of disciplines of authors.}
\label{fig4}
\end{minipage}
\begin{minipage}[t]{0.49\textwidth}
\centering
\caption{Statistics of cross-discipline papers.}
\label{fig5}
\end{minipage}
\end{figure}

There have already been several excellent survey works for the link prediction problem \cite{LK07, HZ11, LZ11}. Liben-Nowell and Kleinberg \cite{LK07} provide useful information and insights for link prediction problem, with special reference to some classical prediction measures based on topological information of networks. This is a pioneering and influential work, but after 7 years, it is not able to cover the latest progress of the link prediction research. 
L\"u and Zhou summarize popular link prediction algorithms for complex networks \cite{LZ11}. However, they emphasize on the contributions from physical perspectives, instead of the perspectives of computer scientists. 
In addition, a complex network is a more abstract model than a real-world social network that we discuss in this paper. 
Therefore, although the link prediction algorithms summarized by L\"u and Zhou are general and valuable, it still needs a new link prediction survey which will focus on social networks.
Hasan and Zaki review some representative link prediction methods for social networks by categorizing them  \cite{HZ11}, especially, the survey mainly considers three types of models proposed in recent years: binary classification model, probabilistic model and linear algebraic model. 
It involves some new representative link prediction works, and it is more suitable for experts, but it is not comprehensive enough for a beginner who wants to learn the link prediction systematically.

To fill up these gaps of existing survey works, this paper tries to provide a comprehensive and systematic survey for link prediction in social networks covering both classical and latest link prediction techniques, link prediction problems, link prediction applications, and active research groups.
First, it gives the link prediction statement including formal definition, general solution framework and evaluation metric, especially, a new category of link prediction is proposed according to two perspectives: link prediction techniques and problems. 
Then link prediction techniques are presented from four kinds of aspects: node-based metrics, topology-based metrics, social-theory-based metrics and learning-based methods. 
Topology-based metrics and learning-based methods contain a lot of classical and new link prediction techniques.  
Based on the link prediction techniques, popular problems that link prediction often faces are discussed. 
Typical link prediction applications are additionally described. 
Active research groups are also presented to demonstrate the different solution perspectives and emphases of leading researchers about link prediction problems.
Finally, besides the current achievements, some future challenges are also discussed.

The structure of this paper is organized as follows. 
In the next section, we present the link prediction definition, general link prediction solution framework, especially the category for link prediction techniques and problems. 
Section 3 will discuss the classical and emerging link prediction techniques. 
Section 4 describes the solutions for different problems in link prediction. 
Section 5 describes applications of link prediction in social networks. 
Section 6 consists of a summary of the work of some active research groups. 
We outline some future challenges in Section 7 and present conclusion in Section 8.

\section{Problem Statement}
Consider a social network $G(V,E)$ at a particular time $t$, where $V$ and $E$ are sets of nodes and links, respectively. The link prediction aims to predict new links or deleted links between nodes for a future time $t'(t'>t)$, or missing links or unobserved links, in current network. This problem can be explained by a simple social network about five persons in Figure 6, in which solid links indicate interactions already existed at time $t$, and dashed links indicate links that have newly appeared during the interval time $[t,t']$. At time $t$, Alice and Bob are friends, Alice and Nick are also friends. At time $t'$, maybe Alice has introduced Bob to Nick, they become friends too. Analogously, Nick and Amy would become friends at time $t'$. The goal of link prediction problem here is to predict the appearance of newly added friendship between persons. 
In this paper, if there are no special declaration, the discussing link prediction work is under an assumption: the nodes in different time are static. Obviously, this assumption does not hold in real-world social networks, and it is one of the challenges in the future.

\begin{figure}[!b]
\centering
\includegraphics[width=0.45\textwidth]{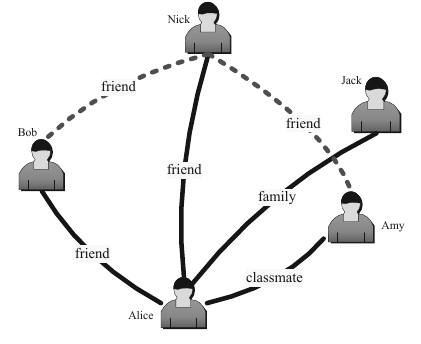}
\caption{An example to explain the link prediction problem.}
\label{fig5}
\end{figure}

\begin{figure}[!t]
\centering
\includegraphics[width=1.0\textwidth]{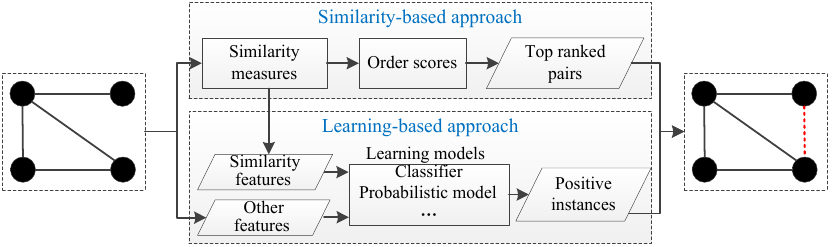}
\caption{The generic link prediction framework.}
\label{fig6}
\end{figure}

To solve the link prediction problem, it needs to determine the formation or dissolution possibilities of links between all node pairs. Usually, such possibilities can be measured by similarities or relative ranks between node pairs. We use a generic framework to illustrate the link prediction solution as shown in Figure 7. For an initial social network, there are two ways to predict the link evolution: \emph{similarity-based approaches} and \emph{learning-based approaches}. 
Here we take predicting the new/missing/unobserved links as examples.
A similarity-based approach is to compute the similarities on non-connected pairs of nodes in a social network, namely, it is based on measures for analyzing the proximity of nodes. 
Every potential node pair $(x, y)$ would be assigned a score, where higher score means higher probability that $x$ and $y$ will be linked in the future, and vice versa. 
Then a ranked list in decreasing order of scores is obtained and links at the top of list are most likely to appear. 
A learning-based approach is treating the link prediction problem as a binary classification task \cite{HCS06}. 
Therefore, some typical machine learning models such as classifier and probabilistic model can be used for solving this problem. 
Each non-connected pair of nodes corresponds to an instance with features describing nodes and the class label. 
If there is a potential link connecting a pair of nodes, this pair is labeled as positive, otherwise it is  negative. 
For the learning-based approaches, the features consist of two parts: one is the similarity features from the similarity-based approaches, another is the features derived from the social network, such as the textual information of attributes and domain knowledge.
Link prediction for deleting/disappearing links can be solved analogously.

There are many link prediction works, that focus on general link prediction techniques,  discuss special link prediction problems, and employ existing link prediction techniques to deal with various applications. For the sake of clearly arranging these existing works, this paper proposes a new link prediction category with two perspectives: technique perspective and problem perspective. 
The category does not contain link prediction applications, since they are based on link prediction techniques and problems, and will be addressed independently. 
Figure 8 shows our category of link prediction techniques and link prediction problems. 
For link prediction techniques, they can be divided into four levels from top to bottom: (1) According to basic network information used in prediction, the first and highest level consists of node, topology, and social theory. (2) In the second level, topology is further divided into neighbor, path, and random walk; social theory is also divided into community, triad, structural hole, tie strength, and homophily. (3) The third level includes popular basic link prediction techniques based on node, neighbor, path, random walk, and social theory. (4) In the fourth level, basic prediction techniques and external information, which includes weights, attributes and knowledge repository, provide features for complicated learning-based techniques, such as feature-based classification, kernel-based learning, probabilistic model and matrix factorization. All link prediction techniques are generic and can be used for solving various link prediction problems and applications. 
From another perspective, link prediction problems are arranged as three levels from bottom to top: (1) According to the object that link prediction problems concern, the first and lowest level divides the link prediction problems into network and link, the former concerns the global characteristic of network and the latter concerns link characteristic of network. (2) In the second level, the network is divided into  heterogeneous network, location network, temporal network, and bipartite network; meanwhile; the link is also divided into  multi-relation links, active/unactive links, and disappearing links. (3) The third level is the link prediction problems that are based on the link prediction techniques; consequently two perspectives join in here. Therefore, two perspectives are closely related. 
Furthermore, we should highlight that the category in Figure 8 can be extended with the progress of link prediction research.

\begin{figure}[!t]
\centering
\includegraphics[width=1.0\textwidth]{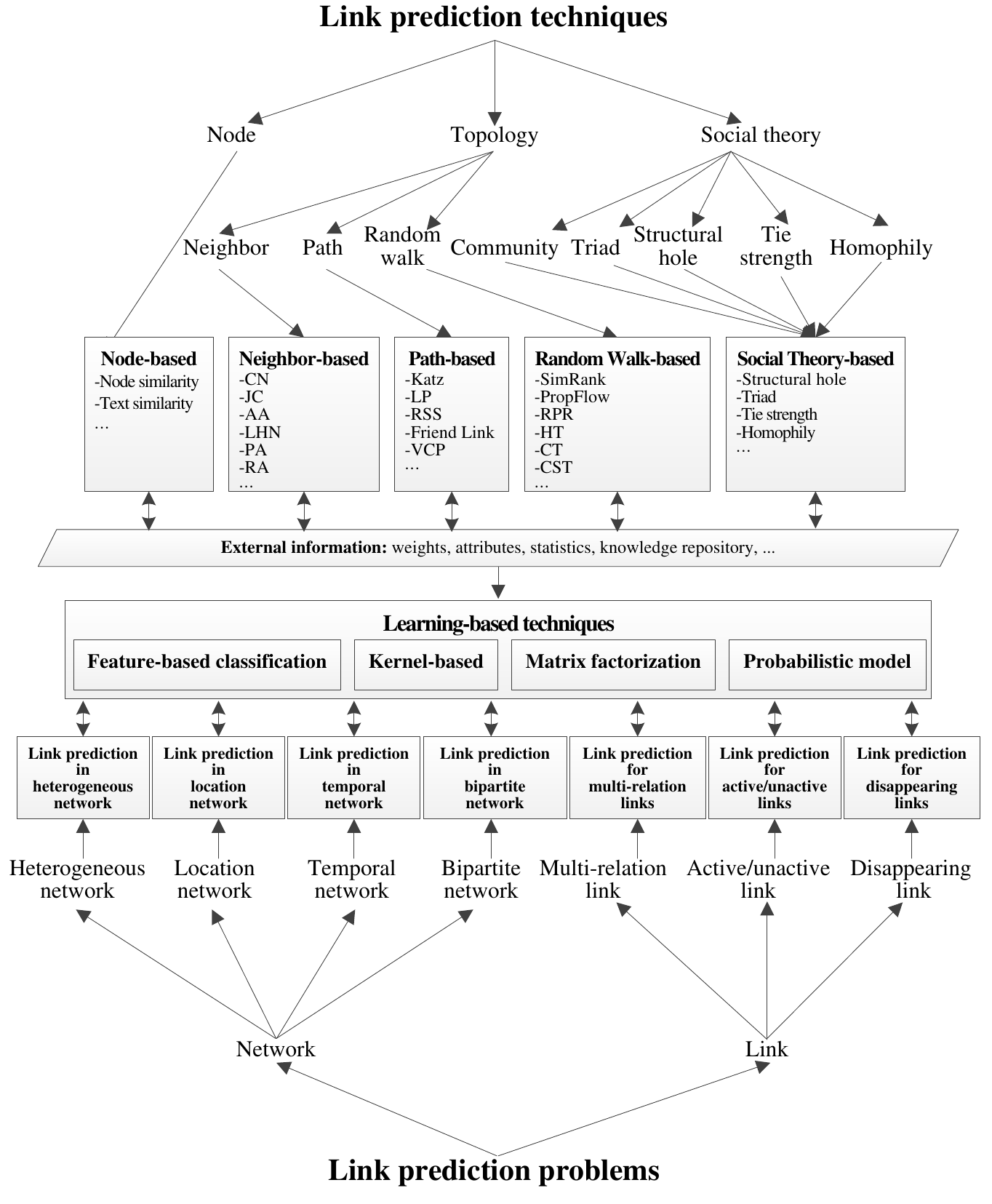}
\caption{The category of link prediction techniques and link prediction problems.}
\end{figure}
    
For the evaluation metrics used in link prediction, they commonly can be divided into two types: fixed threshold metrics and threshold curves. The precision and recall on top-N predictions are typical fixed threshold metrics. These measures often suffer from the limitation that they rely upon a reasonable threshold. Another kind of metrics are the threshold curves, such as the receiver operating characteristic (ROC) curves and precision-recall curves, are widely used in link prediction evaluation. Additionally the AUC (Areas Under ROC) is viewed as a robust measure in the presence of imbalance \cite{HM82}. AUC can be interpreted as the probability that a randomly chosen missing link has higher score than a randomly chosen nonexistent link. It is more difficult to specify and explain link prediction evaluation strategies than with standard classification wherein it is sufficient to fully specify a dataset, therefore, new evaluation methods or performance metrics are also to be proposed \cite{LC12a}.

\section{Link Prediction Techniques}
There are many generic, simple and basic link prediction metrics, which use information of nodes, topology and social theory to calculate the similarities of node pairs. Moreover, learning-based link prediction methods are more complex, but they are established on features provided by the basic metrics and external information. In this section, we will present a systematic review for these link prediction metrics and methods. 

\subsection{Node-based Metrics}

Computing the similarity between a node pair is an intuitive solution for link prediction. It is based on the simple idea: the more similar the pair is, the more likelihood a link between them, and vice versa.
This is consistent with the fact that users tend to create relationships with people who are similar in educations, religions, interests and locations.
It can be measured by the similarity, in which each non-connected pair of nodes $(x, y)$ is assigned a score signifying similarity between $x$ and $y$. A high score indicates high probability that $x$ and $y$ will be linked in the future, while a low score also indicates high probability that $x$ and $y$ will not be linked. Therefore, using the rank of similarity scores, we can predict the appearing or disappearing links in the future or unseen links in current networks.

In a practical social network, a node usually has some attributes such as the profile in online social networks, mail name in email networks, and publication record in academic social networks. 
These information can be directly used for calculating the similarity between two nodes.
Since in most cases the node attribute values are textual forms, the text-based and string-based similarity metrics are usually used here. Discussing classical text-based and string-based similarity measures is beyond the purpose of this paper, and readers can refer to some surveys \cite{HD80, N01a}.

Bhattacharyya et al. define a multiple categorization tree model to study the keywords of user profiles, then they define distance between keywords to determine the similarity between a pair of users \cite{BGW11}. Their most important observation is that except for direct friends, similarity between users are approximately equal, irrespective of the topological metrics. They also show that the increasing of number of friends and keywords lowers the average similarity between the user and his friends.

Akcora et al. found that most user profiles in current social networks are missing. To overcome this limitation in similarity measure, they propose a method to infer a portion of the missing values of a stranger profile before computing similarity\cite{ACF13}. The key idea of this inferring method is based on the profile information of mutual friends and majority voting schema. 

Anderson et al. use users's interests overlap to measure the similarity\cite{AHK12}. Users's interests are represented by the actions they take, such as editing an article on Wikipedia and asking a question on Stack Overflow. All actions of an user can be represented as a vector, then the similarity between two users is the cosine between their respective action vectors.

In conclusion, node-based metrics mainly use the attributes and actions, which can reflect the personal interests and social behaviors, to calculate the similarities between node pairs. Therefore, node-based metrics are useful in link prediction if we can obtain users's attributes and actions in social networks. 

\subsection{Topology-based Metrics}

Even in a simple network without node or edge attributes, there are many metrics available for computing the similarity of two nodes. Most metrics are based on the topological information, and called topology-based metrics. Liben-Nowell and Kleinberg have discussed several metrics based on the graph structural features \cite{LK07}, after their work, many topology-based metrics were proposed. Here, we will give a systematical explanation of popular topology-based metrics in link prediction. According to the characteristics of these metrics, they can be divided into neighbor-based metrics, path-based metrics, and random-walk-based metrics.

For clarity of following descriptions, we first give some standard notations. Let lowercase letters be nodes in the social network and uppercase letters be adjacency matrix of the network. Let matrix $A$ be the adjacency matrix of a given social network. Let $\mit\Gamma(x)$ be the set of neighbors of node $x$, and let $|\mit\Gamma(x)|$ be the number of neighbors of node $x$. 

\subsubsection{Neighbor-based Metrics}

In a social network, people tend to create new relationships with people that are closer to them. Neighbors are the most close ones of a given user. Therefore, researchers design a lot of neighbor-based metrics for link prediction. 

\textbf{Common Neighbors (CN):} The CN metric is one of the most widespread measurements used in link prediction problem mainly due to its simplicity \cite{N01b}. For two nodes, $x$ and $y$, the CN is defined as the number of nodes that both $x$ and $y$ have a direct interaction with. A bigger number of the common neighbors make it easier that a link between $x$ and $y$ will be created. This measure is defined as following formula.
\begin{equation} 
\text{CN}(x,y)=|\mit\Gamma(x)\cap\mit\Gamma(y)|
\end{equation}

Since CN metric is not normalized, it usually reflects the relative similarities between node pairs. Therefore, some neighbor-based metrics consider how to normalize the CN metric reasonably.

\textbf{Jaccard Coefficient (JC):} Jaccard coefficient normalizes the size of common neighbors. It assumes higher values for pairs of nodes which share a higher proportion of common neighbors relative to total number of neighbors they have. This measure is defined as:
\begin{equation} 
\text{JC}(x,y)=\frac{|\mit\Gamma(x)\cap\mit\Gamma(y)|}{|\mit\Gamma(x)\cup\mit\Gamma(y)|}
\end{equation}

\textbf{S{\o}rensen Index (SI):} This metric is defined as formula (3). Besides considering the size of the common neighbors, it also points out that lower degrees of nodes would have higher link likelihood.
\begin{equation} 
\text{SI}(x,y)=\frac{|\mit\Gamma(x)\cap\mit\Gamma(y)|}{|\mit\Gamma(x)| + |\mit\Gamma(y)|}
\end{equation}

\textbf{Salton Cosine Similarity (SC):} SC is a common cosine metric for measuring the similarity between two nodes $x$ and $y$. It is defined as:
\begin{equation} 
\text{SC}(x,y)=\frac{|\mit\Gamma(x)\cap\mit\Gamma(y)|}{\sqrt{|\mit\Gamma(x)| \cdot |\mit\Gamma(y)|}}
\end{equation}

\textbf{Hub Promoted (HP):} HP defines the topological overlap of nodes $x$ and $y$ \cite{RSM02}. It is defined as following formula (5). Obviously, the HP value is determined by the lower degree of nodes.
\begin{equation} 
\text{HP}(x,y)=\frac{|\mit\Gamma(x)\cap\mit\Gamma(y)|}{min(|\mit\Gamma(x)| , |\mit\Gamma(y)|)}
\end{equation}

\textbf{Hub Depressed (HD):} Zhou et al. propose a similar metric to HP \cite{ZLZ09}, but the value is determined by the higher degrees of nodes.
\begin{equation} 
\text{HD}(x,y)=\frac{|\mit\Gamma(x)\cap\mit\Gamma(y)|}{max(|\mit\Gamma(x)| , |\mit\Gamma(y)|)}
\end{equation}

\textbf{Leicht-Holme-Nerman (LHN):} This metric assigns high similarity to node pairs that have many common neighbors compared not to the possible maximum, but to the expected number of such neighbors \cite{LHN06}.
\begin{equation} 
\text{LHN}(x,y)=\frac{|\mit\Gamma(x)\cap\mit\Gamma(y)|}{|\mit\Gamma(x)| \cdot |\mit\Gamma(y)|}
\end{equation}

\textbf{Parameter-Dependent (PD):} To improve the accuracy for predicting both popular and unpopular links, Zhu et al. propose the PD metric as follows \cite{ZLZ12}. Here $\lambda$ is a free parameter. When $\lambda=0$, PD degenerates to CN. If $\lambda=0.5$ and $\lambda=1$, it degenerates to Salton and LHN metric, respectively.
\begin{equation} 
\text{PD}(x,y)=\frac{|\mit\Gamma(x) \cap \mit\Gamma(y)|}{(|\mit\Gamma(x)| \cdot |\mit\Gamma(y)|)^\lambda}
\end{equation}

\textbf{Adamic-Adar Coefficient (AA):} The AA metric was proposed by Adamic and Adar for computing similarity between two web pages at first \cite{AA03}, subsequent to which it has been widely used in social networks. The AA measure is formulated related to Jaccard's coefficient. But here, common neighbors which have fewer neighbors are weighted more heavily. It is defined as:
\begin{equation} 
\text{AA}(x,y)=\sum_{z\in \mit\Gamma(x)\cap\mit\Gamma(y)} \frac{1}{log|\mit\Gamma(z)|}
\end{equation}

\textbf{Preferential Attachment (PA)}: The PA metric indicates that new links will be more likely to connect higher-degree nodes than lower ones \cite{BJN02}. It is defined as:
\begin{equation} 
\text{PA}(x,y)=|\mit\Gamma(x)| \cdot |\mit\Gamma(y)|
\end{equation}

\textbf{Resource Allocation (RA) :} This metric is proposed by Zhou et al. \cite{ZLZ09}, and is motivated by the physical processes of resource allocation. RA metric has a similar form like AA. They both suppress the contribution of the high-degree common neighbors. However, RA metric punishes the high-degree common neighbors more heavily than AA. Therefore, AA and RA have very close prediction results for the networks with small average degrees, but RA performs better for the networks with high average degrees. 
In addition, RA and AA not only use direct neighbors, but also consider neighbors of neighbors. This is different with other metrics.
RA is defined as:
\begin{equation} 
\text{RA}(x,y)=\sum_{z\in \mit\Gamma(x)\cap\mit\Gamma(y)} \frac{1}{|\mit\Gamma(z)|}
\end{equation}

For the reason that neighbors can indirectly reflect users's social behavior and directly affect users's social choice, many link prediction methods are based on neighbors. For example, Akcora et al. propose a new global network similarity \cite{ACF13}, in which they first define the mutual friends graph (MFG) of $x$ and $y$. Edge count of the MFG can measure the strength of ties between $x$ and $y$. To normalize the similarity value, they also define the friendship graph (FG) of node $x$. The network similarity can be measured through a comparison between the number of edges in MFG and the number of edges in FG.

Sarkar et al. presented a theoretical justification of popular link prediction heuristics, to obtain common empirical observations for neighbor-based metrics \cite{SCM11}. They justified that the number of common neighbors gives bounds on similarity of a node pair, therefore, some metrics would predict links with maximum number of common neighbors. They also presented theoretical justification for that metrics carefully using  weighted count of common neighbors that often outperform the unweighted count. Therefore, simple metrics of counting common neighbors often outperform more complicated link prediction methods.

\begin{table}[!t]
\centering
\caption{Comparison of neighbor-based metrics}
\footnotesize

\begin{tabular}{c|c|c|c}
\toprule
  Metric & Normalization & Time complexity & Characteristic\\
  \hline
CN &  No & $O(n^2)$ & Simple and intuitive\\
JC & Yes& $O(2n^2)$ & Proportion of common neighbors relative to total number of neighbors\\
SI & Yes& $O(n^2)$ & Lower degrees of nodes would have bigger link likelihood\\
SC & Yes& $O(n^2)$ & Cosine metric\\
HP & Yes& $O(n^2)$ & Link likelihood is determined by the lower degree of nodes\\
HD & Yes& $O(n^2)$ & Link likelihood is determined by the higher degree of nodes\\
LHN & Yes& $O(n^2)$ & Higher link likelihood to node pairs having many common neighbors\\
PD & Yes& $O(n^2)$ & Improve the accuracy for predicting both popular and unpopular links\\
AA & No& $O(2n^2)$& Common neighbors having fewer neighbors are weighted more heavily\\
PA & No& $O(2n)$& Simple and new links will be more likely to connect higher-degree nodes\\
RA & No& $O(2n^2)$& Similar to AA, but punishes high-degree common neighbors more heavily\\
\bottomrule
\end{tabular}
\end{table}

In Table 2, we compare the popular neighbor-based metrics in normalization, time complexity and characteristic. Four metrics: CN, AA, PA and RA, are not normalized, that means the similarities under these metrics only have the ranking meanings and are not easy be assembled with other normalized metrics. Time complexity is also an important factor when we select metrics, especially for large scale social networks. Assume that the average number of neighbors in a network is $n$, for two nodes $x$ and $y$, the time complexity of finding all neighbors of a node is $O(n)$, and the time complexity of calculating the intersection or union of two sets is $O(n^2)$. CN, SI, SC, HP, HD, LHN, PD have $O(n^2)$ time complexity because that they need to calculate an intersection of two sets. JC's time complexity is $O(2n^2)$ because it calculates an intersection and a union of two sets. AA and RA need to calculate an intersection of two sets and find neighbors of common neighbors, therefore, their time complexities are $O(2n^2)$. PA only needs to find neighbors of $x$ and $y$, and its time complexity is $O(2n)$. Characteristics of neighbor-based metrics are also listed as discussed above. This comparison can help people to choose suitable metrics for practical social networks.

Finally, we should draw attention to the fact that, although there are many existing neighbor-based metrics, but in practical applications, one should choose right metrics according to the characteristics of social networks,
because many experiment evaluation results have shown that there is no an absolutely dominating metric for different datasets \cite{LK07}.

\subsubsection{	Path-based Metrics}

Besides node and neighbor's information, paths between two nodes can also be used for computing similarities of node pairs, and we call such methods path-based metrics. 

\textbf{Local Path (LP):} LP metric \cite{LJZ09} makes use of information of local paths with length 2 and length 3. Unlike the metrics that only use the information of the nearest neighbors, it exploits some additional information of the neighbors within length 3 distances to current node. Obviously, paths of length 2 are more relevant than paths of length 3, so there is an adjustment factor $\alpha$ applied in the measure. $\alpha$ should be a small number close to 0. The metric is defined as following formula (12). Here, $A^2$ and $A^3$ denote adjacency matrices about the nodes having 2 length and 3 length distances, respectively. Therefore, LP is also an adjacency matrix which describes the node pairs with length 2 and 3 distances.
\begin{equation} 
\text{LP} = A^2 + \alpha A^3
\end{equation}

\textbf{Katz:} Katz metric \cite{K53} is based on the ensemble of all paths, and it counts all paths between two nodes. The paths are exponential damped by length that can give more weights to the shorter paths. This measure is defined as follows, where $path^{l}_{x,y}$ is the set of all paths from $x$ to $y$ with length $l$, $\beta>0$ and the very small $\beta$ will cause Katz metric much like CN metric because paths of long length contribute very little to final similarities.
\begin{equation} 
\text{Katz}(x,y)=\sum_{l=1}^{\infty} \beta ^l \cdot |path^{l}_{x,y}| = \beta A+\beta ^2 A^2 + \beta ^3 A^3 + \cdots
\end{equation}

\textbf{Relation Strength Similarity (RSS):} RSS \cite{CGZ12A} is an asymmetric metric that can be used for the weighted social networks. It is calculated based on relation strength $R(x, y)$, a normalized link weighting score defining the relative degree of similarity between neighboring nodes. Assuming that there are $L$ simple paths $p_1$, $p_2$,$\cdots$, $p_L$ shorter than $r$ from $x$ to $y$, and path $p_l$ is formed by $K$ nodes $z_1$, $z_2$, $\cdots$, $z_{k-1}$ and $z_k$. Then the RSS from $x$ to $y$ is defined as:
\begin{equation} 
\text{RSS}(x,y)=\sum_{l=1}^{L}R_{p_l}^*(x,y)
\end{equation}
\begin{equation} 
R_{p_l}^*(x,y)=\left\{
\begin{array}{ll}
\prod_{k=1}^{K}R(z_k,z_{k+1}) & K\le r \\
0 & otherwise
\end{array}
\right.
\end{equation} 

\textbf{FriendLink (FL):} FL metric \cite{PSM12} is a similarity between node $x$ and $y$, by traversing all paths of a bounded length. It can provide more accurate and faster link prediction. FL assumes that persons in a social network can use all the paths between them, proportionally to the path lengths. The similarity between $x$ and $y$ is defined as counts of paths of varying length l from $x$ to $y$:
\begin{equation}
\text{FL}(x,y)=\sum_{i=1}^l \frac{1}{i-1} \cdot \frac{|paths_{x,y}^i|}{\prod_{j=2}^{i}(n-j)}
\end{equation}
where $n$ is the number of vertices in network, $l$ is the length of a path between $x$ and $y$ (excluding path with cycles), $paths_{x,y}^i$ is the set of all $length-i$ paths from $x$ to $y$. However, it does not mean a higher $l$ will produce higher precision. In fact, precision will follow a degraded performance for higher $l$.

\textbf{Vertex Collocation Profile (VCP):} VCP is proposed for link analysis and prediction based on a restrictive representation of the local topological embedding of two nodes\cite{LC12b, LC14}. Formally, a VCP, written as $\text{VCP}_{x,y}^{n,r}$, is a vector describing the relationship between two nodes $x$ and $y$, in terms of their common membership in all possible subgraphs of $n$ vertices over $r$ relations.
Obviously, VCP cannot produce similarities between nodes, but it can be the classification feature vector for supervised learning methods that will be discussed in later section. For the reason that a subgraph in VCP can be seen as the combination of multiple paths, we categorize VCP as a path-based metric. 

Different from the link prediction models compressing a selection of simple information in theoretically or empirically guided ways, VCP approach preserves as much topological information as possible about the embedding of node pairs. It also extends naturally to multi-relational networks and can thereby encode a variety of additional information such as edge directionality. It even can encode continuous quantities such as edge weights by binning into relational categories, such as high activity and low activity. However, the number of subgraphs of VCP depends on $n$ and $r$, that grows exponentially. For example, $\text{VCP}^{4,1}$ has 32 subgraphs and  $\text{VCP}^{5,2}$ has 524,288 subgraphs. Therefore, it would fail to return results within a reasonable time for large scale networks.

Some link prediction methods use local network information, and others use global network information. The former has computing superiority, and the latter exhibits higher accuracy but cannot handle the large networks due to high time complexity. For example, to maintain relatively high accuracy and simultaneously to take less computing effort, Hu et al. proposed a trade-off method called semi-local similarity index by introducing the resource allocation process into the local path index \cite{BHT11}.

Feng et al. investigated the impact of the network structure on the performance of link prediction methods in the view of clustering \cite{FZX12}. The experimental results show that as the clustering grows, the precision of link prediction could be improved remarkably, while for the sparse and weakly clustered network, they perform poorly. This phenomenon is due to the distinguishment between the distribution of positive and negative instances caused by the variation of clustering. This would be helpful for choosing right link prediction methods when we meet real world networks with various clusterings.

Compared with node-based and neighbor-based metrics, which only use local topological information, path-based metrics consider more topological information: not only local neighbors, but also a kind of important global information, namely, paths between node pairs. Time complexity of path-based metrics is higher than neighbor-based ones. However, longer paths are not always more useful than shorter paths. 
In the theoretical justification of popular link prediction heuristics by Sarkar et al. \cite{SCM11}, it shows that metrics considering longer paths is  useful only if shorter paths are not numerous enough. Therefore, path-based metrics can optimize their performances by removing too long paths if networks have enough shorter paths.

\subsubsection{Random Walk based Metrics}

Social interactions between nodes in social networks can also be modeled by random walk, which uses transition probabilities from a node to its neighbors to denote the destination of a random walker from current node. 
There exists some link prediction metrics which calculate similarities between nodes based on random walk. 

\textbf{Hitting Time (HT):} 
$\text{HT}(x,y)$ is the expected number of steps required for a random walk from node $x$ to node $y$. 
Let $P=D_{A}^{-1}A$, where diagonal matrix $D_{A}$ of $A$ has value $(D_{A})_{i,i}=\sum_{j}A_{i,j}$ and $P_{i,j}$ is the probability of stepping on node $j$ from node $i$.
It is defined as follows \cite{FPR07}:
\begin{equation} 
\text{HT}(x,y)=1+ \sum_{\omega\in \mit\Gamma(x)}P_{x,\omega}\text{HT}(\omega,y)
\end{equation}

\textbf{Commute Time (CT):}
Since the hitting time metric is not symmetric, commute time is used to count the expected steps both from $x$ to $y$ and from $y$ to $x$. It can be obtained as follows:
 \begin{equation} 
\text{CT}(x,y)=\text{HT}(x,y)+\text{HT}(y,x)=m(L_{x,x}^{\dagger}+L_{y,y}^{\dagger}-2L_{x,y}^{\dagger})
\end{equation} 
where $L^\dagger$ is the pseudo-inverse of matrix $L=D_A-A$, $m$ is the number of edges in a social network.

\textbf{Cosine Similarity Time (CST):}
The cosine similarity time metric is based on $L^\dagger$ by calculating similarity of two vectors, and it can be defined as follows:
\begin{equation} 
\text{CST}(x,y)=\frac{L_{x,y}^{\dagger}}{\sqrt{L_{x,x}^{\dagger}L_{y,y}^{\dagger}}}
\end{equation} 

\textbf{SimRank:} SimRank metric \cite{JW02} is defined in a self-consistent way, according to the assumption that two nodes are similar if they are connected to similar nodes. There is a parameter $\gamma$ that controls how fast the weight of connected nodes decrease as they get farther away from the original nodes.
\begin{equation}
\text{simRank}(x,y)=\left\{
\begin{array}{ll}
1 & x = y \\
\gamma \cdot \frac{\sum_{a\in \mit\Gamma(x)} \sum_{b\in \mit\Gamma(y)} simRank(a,b)}{| \mit\Gamma(x)|\cdot| \mit\Gamma(y)|} & otherwise
\end{array}
\right.
\end{equation}
The SimRank score can be explained in terms of the random surfer-pairs model: $simRank(x, y)$ measures how soon two random surfers are expected to meet at the same vertex if they individually start at vertices $x$ and $y$, and randomly walk through edges on the reverse graph. 

The computation complexity of SimRank is $O(n^4)$ at the worst time where $n$ is the number of vertices. Such a high computation cost limits its wide usage for large scale networks.

\textbf{Rooted PageRank (RPR):} Rooted PageRank \cite{LK07} is a modification of PageRank, which is the core algorithm used by search engine to rank search results. The rank of a node in graph is proportional to the probability that the node will be reached through a random walk on the graph. In addition, there is a factor $\epsilon$ that specifies how likely the algorithm is to visit the node's neighbors than starting over. Let $D$ be a diagonal matrix with $D_{i,i}=\sum_jA_{i,j}$. The measure is defined as:
\begin{equation}
\text{RPR} =(1-\epsilon)(I-\epsilon D^{-1} A)^{-1}
\end{equation}

\textbf{PropFlow:} PropFlow metric \cite{LLC13} is similar to Rooted PageRank, but it is more localized. It is proportional to the probability that a restricted random walk starting at $x$ and ending at $y$ in no more than $l$ steps. The restricted walk selects links based on weights and terminates when it reaches $y$ or revisits any nodes. This produces a score that can serve as an estimation of the probability of new links.

If $x$ and $y$ are directly linked, their PropFlow $\text{PF}(x,y)$ can be calculated by follows:
\begin{equation}
\text{PF}(x,y) =\text{PF}(a,x)\frac{w_{xy}}{\sum_{k\in \mit\Gamma(x)}w_{xk}}
\end{equation}
where $k$ is $x$'s neighbor whose depth is greater than the depth of $x$ from the staring node, $w_{xy}$ denotes the weight of the link between nodes $x$ and $y$, and $a$ is the previous node of $x$ on a random walk path. If $x$ is the staring node, $\text{PF}(a,x)=1$. 

If $x$ and $y$ are indirectly linked, $\text{PF}(x,y)$ is the sum of PropFlow through all the shortest paths from $x$ to $y$. 

Unlike rooted PageRank, the computation of PropFlow does not require walk restarts or convergence but simply employs a modified breadth-first search restricted to height $l$. Therefore, it is a faster metric than rooted PageRank and simRank.

Symeonidis et al. proposed SepctralLink, which enhances the multi-way spectral clustering method by introducing new ways to capture node proximity \cite{SIM13}. The new enhanced method uses information obtained from the top few eigenvectors of the normalized Laplacian matrix. As a result, it produces a less noisy matrix, which is smaller and more compact than the original one. In this way, it can provide faster and more accurate link predictions. Moreover, this model is based on the well-known Bray–Curtis coefficient to measure proximity between two nodes. Compared to traditional clustering algorithms, which assume globular (convex) regions in Euclidean space, this approach is more flexible in capturing the non-connected components of a social graph and a wider range of cluster geometries. Symeonidis et al. also extend SepctralLink for social networks with positive and negative links \cite{SM13}.  Compared to metrics based on local network information such as neighbors and paths with length 2, the SepctralLink captures the global network structure by exploiting the normalized Laplacian matrix of the graph, thus improving the prediction accuracy. Compared to metrics based on global network information such as random walk, the SepctralLink is more efficient since that it is based on the top few eigenvectors and eigenvalues of the normalized Laplacian matrix, and needs less time and space complexity than global metrics such as Katz and SimRank.

\subsection{Social Theory based Metrics}

In recent years, more and more works have employed classical social theories, such as community, triadic closure, strong and weak ties, homophily, and structural  balance,  to solve the social network mining and analyzing problems. Different to previous metrics, which only use node and topology, the link prediction metrics based on social theory can improve the performance by capturing useful additional social interaction information, especially for large scale social networks. 

Valverde-Rebaza and Lopes combined topology with community information by considering users's interest and behaviors, then predict future links in Twitter \cite{VL13}. It shows that this method can efficiently improve the link prediction performance in the directed and asymmetric large scale social networks.

Liu et al. proposed a link prediction model based on weak ties and three node centralities of common neighbors: degree, closeness and betweeness centrality \cite{LHH13}. Each common neighbor plays a different role to the node connection likelihood according to their centralities. The weak tie is considered for improving the prediction accuracy. This model can be defined as follows:
\begin{equation}
\text{LCW}(x,y)=\sum_{z}(w(z)\cdot f(z))^{\beta}, \qquad f(z)=\left\{
\begin{array}{ll}
1 & z \in \mit\Gamma(x) \cap \mit\Gamma(y) \\
0 & otherwise
\end{array}
\right.
\end{equation}
where $w(z)$ denotes the weight of node centrality, $f(z)$ is the switch function, and $\beta$ can adjust the contributions of each common neighbor to the connection likelihood of two nodes. When $\beta$ is greater than 1, it makes the larger centrality more significant than the lower centrality. When $\beta$ is less than 0, it restrains the large centrality more than the lower centrality. When $\beta$ is in range (0, 1), it equally restrains all nodes.

Li et al. pointed out that the centrality of nodes is also important for link prediction, namely, nodes in a network not only prefer to link to similar nodes, but also prefer to link to the central nodes. 
They propose a set of link prediction methods based on maximal entropy random walk, which can capture the centrality of nodes \cite{LYL11}. These new methods outperform all the older ones without centrality.

Qiu et al. proposed a behavior evolution-aware event-driven locality and attachedness based model to capture the growth dynamics in social networks \cite{{QIY10}}. This model can better characterize the growing process and simulate important network structures observed in real networks. Then node behavior is used to improve the link prediction accuracy \cite{QHY11}. 
 
In interest networks, homophily is also exploited to predict not only links between a user and his interested services, but also links between users who have common interests \cite{YLS11}. 
In addition, degree distribution, social balance and microscopic mechanism are useful for finding social patterns across different social networks \cite{DTW12}. These patterns can be used to devise complicated link prediction methods.

\subsection{Learning-based Methods}

Based on the features provided by previous basic link prediction metrics, internal attributes, and external information, many learning-based link prediction methods are proposed in recent years. These learning-based methods can be divided into feature-based classification, probabilistic graph model and matrix factorization.

\subsubsection{Feature-based Classification}
Let $x,y\in V$ be nodes in the graph $G(V,E)$ and  $l^{(x,y)}$ be the label of the node pair instance $(x, y)$. In link prediction, each non-connected pair of nodes corresponds to an instance includes the class label and features describing the pair of nodes. Therefore, a pair of nodes can be labeled as positive if there is a link connecting the nodes, otherwise, the pair is labeled as negative. The label of $x$ and $y$ is defined as follows: 
\begin{equation}
l^{(x,y)}=\left\{
\begin{array}{ll}
+1 & if (x,y) \in E \\
-1 & if (x,y) \notin E
\end{array}
\right.
\end{equation}
This is a typical binary classification problem and many supervised classification learning models can be used to solve it. For instance, decision tree, support vector machines, na\"ive Bayes, and so on. 

In order to build an efficient classifier for link prediction, it is crucial to define and extract a set of appropriate features from social networks. 
The features provided by node-based, topology-based and social theory based metrics are popular and important for classification learning models. 
For example, the VCP metric can be seen as a kind of special feature which describes local topology information \cite{LC12b}. 
In addition, many studies show that using attributes of nodes and links  (such as users' ages, interests, characteristics and friends) can significantly improve the link prediction performance. 
Li et al.  presented a graph kernel-based learning method and used features such as age, education level, book title, keywords and introduction to predict user-item link in a bipartite network \cite{LC13}. 
Scellato et al.  considered social features, place features and global features in location-based social networks for link prediction based on a supervised learning framework \cite{SNM11}. 
The co-authorship social network is one of the most popular networks in link prediction. 
Ichise et al. \cite{PI07,WI08} introduced a semantic approach using abstract information, research titles and event information to improve the link prediction in a  co-authorship social network.
Table 3 shows some commonly used features in a co-authorship network, and it contains node features, network features, topological features and non-topological features. 
The advantage of non-topological features is that it can improve the performance of link prediction problem. However, they are not always available and may be difficult to collect. 
More importantly, most non-topological features are domain related. It requires good domain knowledge to identify and discover them. Therefore, for a general link prediction learning model, it usually only considers the generic features including node, network and topological features, but for a practical link prediction application, non-topological features should also be considered.    
Finally, like many learning problems, feature selection in link prediction learning is the key task. Fortunately, a lot of feature selection work in machine learning area can be used. For example, 
Scripps et al. proposed a discriminative learning technique for link prediction based on the matrix alignment\cite{STC08}. The advantage of this method is that it can automatically determine the most predictive attributes and topological features by aligning the adjacency matrix of a network with weighted similarity matrices computed from node attributes and neighborhood topological features .

\begin{table}[!b]
\centering
\caption{Common features in a co-authorship network}
\footnotesize

\begin{tabular}{c|p{320pt}}
\toprule
  Type &Feature\\
  \hline
\multirow{1}{*}{Node} & in degree, out degree\\	
\hline
\multirow{2}{*}{Network} & number of vertices, number of edges, average clustering coefficient\\
	& mean degree, number of strongly-connected components(SCC), largest SCC \\
\hline
\multirow{3}{*}{Topological}	
	& common neighbors, Jaccard Coefficient, Adamic-Adar, Preferential Attachment\\
	& Resource Allocation, Local path, Katz, Rooted PageRank, SimRank\\
	& PropFlow, Relation Strength Similarity, Max Flow, shortest paths\\
\hline
\multirow{3}{*}{Non-topological} 
	 & co-authors, number of neighbors, keyword counts, keywords match count\\
	 & clustering index, community alignment, publish date\\
	 & paper title, author affiliation, paper venue\\
\bottomrule
\end{tabular}
\end{table}

Usually, people would think that the weight is also an important feature for link prediction learning. However, the relevance of weights for unsupervised link prediction methods is not always verified, in some cases, even the performance is  harmed. For supervised link prediction methods, there are few works show that weights would improve the prediction results \cite{DP11}. But there are also some works believe that weights are useless for unsupervised link prediction \cite{LZ09}. Therefore, it still needs comprehensive analysis on more datasets and applications in theory and evaluation study to determine the influence of weights.

Based on the transformation of a graph's algebraic spectrum, Kunegis and Lommatzsch presented a unified framework for learning both link prediction and edge weight prediction functions\cite{JL09}. 
This framework generalizes graph kernels and dimensionality reduction methods and estimates the parameters efficiently. 
First it derives variants that apply to undirected, weighted, unweighted, unipartite and bipartite graphs, then generalizes existing link prediction functions to a common form.
And then it provides a way to reduce the high-dimensional problem of learning the various kernels's parameters to a one-dimensional curve fitting problem.

Pujari and Kanawati proposed a new dyadic topological link prediction approach applying supervised social choice algorithm \cite{PK12}. They used these data to learn weights to associate to each computed feature based on the ability of each attribute to predict observed links. These weights were then used within weighted/supervised computational social choice algorithms to predict new links. 
They introduced weighted social choice rules by modifying classical voting approaches, 
and then combined the prediction power of individual topological measures by applying computational social choice algorithms (or what is also known as rank aggregation methods). Rank aggregation can be defined as a process of combining a number of ranked lists or rankings of candidates or elements to get a single list and with least possible disagreement with the all the experts or voters who provide these lists. 
It is a part of social choice theory and has been applied to political and election related problems. Expressing the link prediction problem in terms of a vote is straightforward: candidates are examples (pairs of unconnected nodes), while voters are topological measures computed for these pairs of unlinked nodes.

In signed networks, signed links reflect social attitudes of the users towards each other such as friendship, hostile or trust.
Wang et al. showed that social imbalance in a network can be used to derive a link prediction method \cite{CNT11}. 
A supervised learning based link prediction method is proposed and uses features derived from longer cycles in a network. 
After investigating a kind of signed networks, whose relationships can be either positive (indicating relations such as friendship) or negative (indicating relations such as opposition or antagonism), Leskovec et al. found that the signs of links in such social networks can be predicted with high accuracy by using models that provide insight into some of the fundamental principles that drive the formation of signed links \cite{LHK10}.
It is consistent with social balance theory and social psychology status.
It also means that social attitude of one user toward another can be estimated from evidence provided by their relationships with other members of the surrounding social network.
  
Cao et al. discussed the data sparsity problem in link prediction by jointly considering multiple link prediction tasks from heterogeneous domains such as predicting links between users and different types of items, which is referred to as the collective link prediction problem \cite{CLY10}. They proposed a transfer learning idea to solve the problem by a nonparametric Bayesian framework, which allows knowledge to be adaptively transferred across heterogeneous tasks while taking into account the similarities between tasks. 

To utilize the auxiliary social networks or available proximity networks, Lu et al. proposed a supervised learning framework that can effectively and efficiently learn the dynamics of social networks in the presence of auxiliary networks, then construct a rich variety of path-based features using multiple source for link prediction \cite{LST10}. 

Wu et al. proposed an interactive learning framework to formulate link prediction into a factor graph model  \cite{WST13}. As Figure 9 shows, it consists of three stages : (1) Perform similarity measures to generate candidate nodes based on various features such as homophily, referral chaining and recency.
(2) Perform a ranking factor graph (RankFG) model to refine the ranking. The model integrates two types of factor functions, pairwise factor function and correlation factor function, to capture characteristics of node pairs and correlations between suggested results.
(3) User can provide feedback to the suggested relationships. 
An interactive learning algorithm RankFG+ is designed to adjust the ranking model incrementally based on the user's feedback.

\begin{figure}[!htb]
\centering
\includegraphics[width=0.9\textwidth]{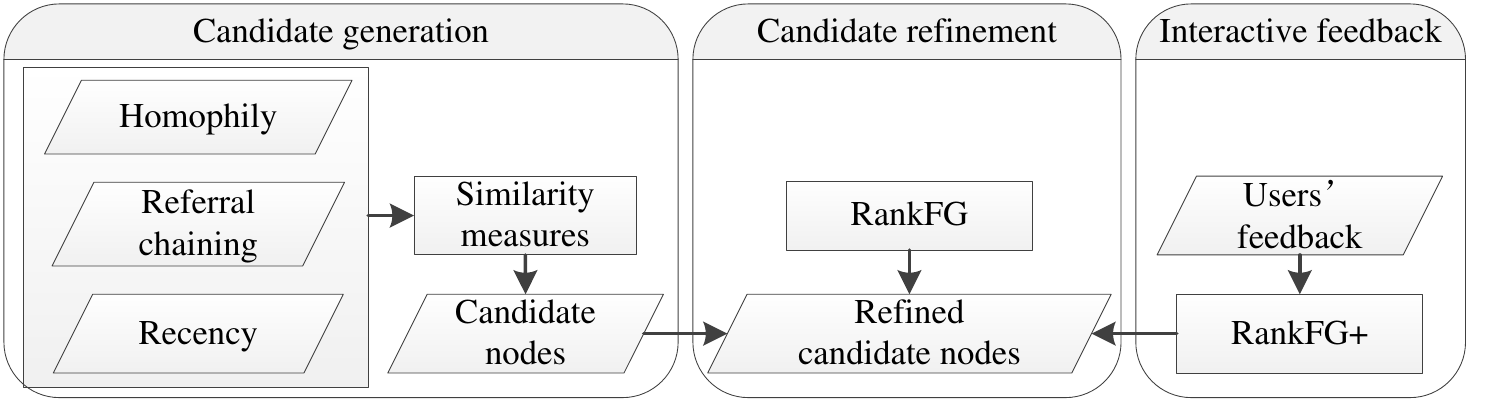}
\caption{Three stages of the interactive learning framework.}
\label{fig9}
\end{figure}

Besides the supervised classification model, the semi-supervised models can also be used to solve the link prediction problem. 

Kashima et al. proposed link propagation as a new semi-supervised learning method for link prediction problems \cite{KKY09}, where the task was to predict unknown parts of the network structure by using auxiliary information such as node similarities. Since the proposed method can fill in missing parts of tensors, it is applicable to multi-relational domains, allowing us to handle multiple types of links simultaneously. 
This method applies the label propagation method to link prediction, and it is the first method for tensor completion using auxiliary information. It can handle not only strength of links among pairs of nodes, but also various types of links. 
Despite its efficiency and effectiveness compared to other methods, its applications were still limited due to the computational time and space constraints. A fast and scalable algorithm is proposed for the Link Propagation by introducing efficient procedures to solve large linear equations that appear in the method  \cite{RK10}. In particular, it shows how to obtain a compact representation of the solution to the linear equations by using a non-trivial combination of techniques in linear algebra to construct algorithms that are also effective for link prediction on dynamic graphs. 
Brouard et al. also addressed link prediction as an output kernel learning task through semi-supervised output kernel regression \cite{BS11}. 

By comparing some feature-based learning methods as shown in Table 4, we can obtain following observations: 
(1) Link prediction methods usually directly employ basic classification models or modify other well-known models \cite{KKY09}. It means that classification models are not the key points in link prediction learning.
(2) Actually, feature selection or construction is the core task in link prediction learning, and it is the main difference between these link prediction learning methods. The general learning frameworks mainly use topological and node features, but for handling domain-specific link prediction problems, domain-specific features should be constructed and used \cite{WST13, SNM11}.
(3)The supervised link prediction methods can improve the prediction performance, especially the prediction precision. Meanwhile, it would also cause the high computing cost in feature selection and model training \cite{LC12b, PK12}. 

\begin{table}[]
\centering
\caption{Comparison of feature-based classification methods}
\footnotesize

\begin{tabular}{p{55pt}|p{85pt}|p{85pt}|p{70pt}|p{100pt}}
\toprule
  Methods & Features & Learning models & Network types & Strength or weakness\\
  \hline
 VCP\cite{LC12b} & Vertex collocation profile &  Classification models in WEKA, HPLP supervised link prediction framework & Directed, weighted, temporal, multi-relational networks & Preserves as much topological information as possible; low performance in featuring and training\\
 \hline
 Generic kernel-based machine learning\cite{LC13} &Topological features: random walk paths; node features: reflecting users' decisions   & One-class SVM kernel machine  & User-item bipartite network &Do not require explicit feature generation; performance is highly dependent on kernel functions\\
\hline
Scellato et al. \cite{SNM11} & Place features, social features, global features & Classifiers in WEKA: J48, Na\"ive Bayes, model trees, random forests& Location-based social networks & Achieve high precision\\
\hline
Ichise et al. \cite{PI07,WI08} & Topological features, semantic and event-based features & 
SVM, desision trees, J48, decision stump, boosting & Co-authorship networks& Performance is dependent on classification models\\
\hline
Scripps et al.\cite{STC08} & Node attributes, neighborhood topological features & Discriminative classifiers& General networks& Flexible framework that automatically determines the most predictive features\\
\hline
S\'a and Prud\^encio \cite{DP11} &Features of topology-based metrics &J48, Na\"ive Bayes, IBk, libSVM, LibLinear&Weighted networks&Using weights can improve supervised link prediction\\
\hline
Spectral graph transformations \cite{JL09}&Adjacency matrix, number and length of paths&Laplacian Kernels& Undirected, weighted, unweighted, unipartite and bipartite large networks&General graph kernels and dimensionality reduction; runtime only depends on the chosen reduced rank, and is independent of the original graph size \\
\hline
Pujari and Kanawati \cite{PK12} &Topological features&Supervised rank aggregation, decision tree, na\"ive Bayes, kNN&General networks&Aggregate features as many as possible; high  time complexity.\\
\hline
Chiang et al.\cite{CNT11}&Features from longer cycles&Logistic regression&Signed networks&Not only achieves good accuracy for sign prediction but is also effective in lowering the false positive rate\\
\hline
Leskovec et al. \cite{LHK10}&Degrees of the nodes, triad&Logistic regression&Signed networks&Signs of links can be predicted with high accuracy\\
\hline
Collective link prediction\cite{CLY10}&No specific features restricted&Nonparametric Bayesian model, transfer learning&Large user-item networks&Transfer learning could help boost the performance\\
\hline
Penalized output kernel regression\cite{BS11}&No specific features restricted&Output kernel regression&General networks& Uses the unlabeled data to improve performances for a low percentage of known links\\
\hline
Lu et al.\cite{LST10}& Path-based features& Logistic regression& Social networks with multiple auxiliary networks& Prediction accuracy is improved; do not consider other features\\
\hline
Wu et al. \cite{WST13}& Statistics of co-inventors, link homophily, interest homophily, and correlation& Ranking factor graph model; interactive learning&Enterprise social networks& Significantly improves the performance for recommending co-invention relationships\\
\hline
Link propagation \cite{KKY09}&Node features: Kronecker sum similarity, Kronecker product similarity&Link propagation&Multi-relational networks&Handles strength and types of links; high computational time and space\\
\hline
Fast and scalable link propagation \cite{RK10} & Node features: Kronecker sum and product&Link propagation;  matrix factorization and approximation & Large dynamic networks &Less computational cost while maintaining accuracy\\

\bottomrule
\end{tabular}
\end{table}

\subsubsection{Probabilistic Graph Model}
In a social network, a link between each node pair can be assigned a probability value such as a topological similarity or transition probability in random walk. 
It is a probabilistic graph. 
There are many learning-based link prediction methods that have been proposed by exploiting the probabilistic graph model. 

Studies suggest that many networks exhibit hierarchical structure, where nodes divide into groups that can further subdivide into groups of groups, and so forth over multiple scale. Clauset et al.\cite{CCN08} proposed a model to infer hierarchical structure from network and apply it to solve the link prediction problem. As shown in Figure 10, a hierarchical network is represented by a dendrogram called hierarchical random graph, where $N$ leaves corresponding to nodes of network and each of the $N-1$ internal nodes corresponds to a probability $p_r$. The connecting probability of a pair of nodes equals to $p_r$ where $r$ is the lowest common ancestor of the two nodes. For example, according to the dendrogram in Figure 10, the connecting probability of node $a$ and node $c$ is 0.6, and the connecting probability of node $b$ and node $d$ is 0.3. Then the goal is to find the dendrogram that fits the observed network best.
Given a network $G$ and a dendrogram $D$, let $E_r$ be the number of edges where $r$ is the lowest common ancestor of two nodes, and let $L_r$ and $R_r$ be the numbers of leaves in the left and right subtrees rooted at $r$. Then the likelihood of the network is:
\begin{equation}
 L(D,\{P_r \})=\prod_rp_r^{E_r}(1-p_r)^{L_rR_r-E_r}
\end{equation}
The probability of internal nodes is easy to determine by maximizing $L(D,\{P_r\})$.
To predict whether a pair of non-connected nodes $x$ and $y$ are connected, it first samples a set of dendrograms with probability proportional to their likelihood, then computes the mean probability $p_{xy}$ over the sample dendrograms by averaging the corresponding probability $p_{xy}$.
Compared to the basic metrics such as common neighbor, the hierarchical random graph model is able to express assortative and disassortative structure and obtain accurate predictions for a wide range of networks. However, it takes a lot of computing time and usually is applied to deal with networks within thousands of nodes.

\begin{figure}[!htb]
\centering
\includegraphics[width=0.6\textwidth]{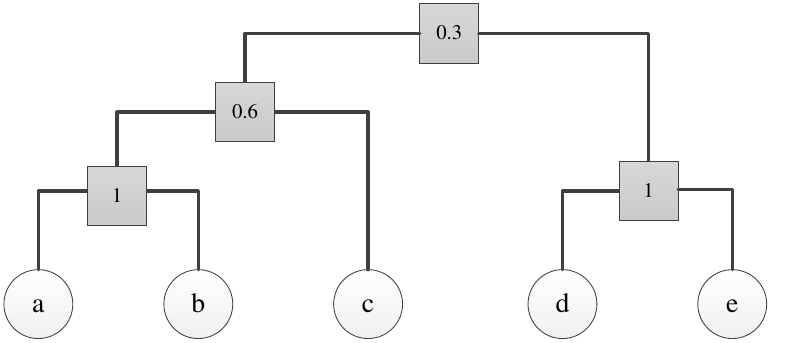}
\caption{A dendrogram of a network.}
\label{fig10}
\end{figure}

In stochastic block model, nodes in the network are divided into several groups and nodes in the same group have same status \cite{GS09}. That is to say, whether two nodes will connect depends on the group to which they belong. A stochastic block model $M = (P, Q)$ consists of two parts: a method of partition $P$ and a probability matrix $Q$ of two nodes in two different groups. Let $Q_{\alpha\beta}$ be the connecting probability between group $\alpha$ and group $\beta$, and $A$ be the observed network. Then the likelihood of the network structure is:
\begin{equation}
p(A|P,Q)=\prod_{\alpha \le \beta}Q_{\alpha\beta}^{l_{\alpha\beta}}(1-Q_{\alpha\beta})^{\gamma_{\alpha\beta}-l_{\alpha\beta}}
\end{equation}
where $l_{\alpha\beta}$ is the number of the existing links between group $\alpha$ and $\beta$ and $r_{\alpha\beta}$ is the number of possible maximum links between group $\alpha$ and $\beta$.
Let $\Omega$ be the set of all possible partitions. The reliability of a particular link $(x, y)$ can be calculated according to the Bayes' Theorem:
\begin{equation}
p(A_{xy=1}=1|A)=\frac{1}{z}\sum_{p\in\Omega}\int_0^1|Q|p(A_{xy=1}=1|P,Q)p(A|P,Q)p(P,Q)dQ
\end{equation}
where $Z$ is a normalization factor which can be represented as $Z=\sum_{p\in\Omega}exp[-H(P)]$, and $H(P)$ is a function of the partition.
This model is able to identify both missing and spurious links in noisy network observations and renders better prediction than hierarchical random graph model. 
However, it also takes huge computation time and lacks the ability to capture the possible overlapping or hierarchical structure. The marginalized denoising model proposed by Chen and Zhang could overcome this shortcomings \cite{CZ14}. It treats the link prediction as a problem of matrix denoising, and the key point of this method is learning a mapping function that can map the  matrix of current network with observed links to a new matrix of future network with unobserved links. 

Wang et al. proposed a method which utilized three types of features, namely, co-occurrence probability features, topology features and semantic features, to solve the link prediction problem \cite{WSP07}. This method is described in Figure 11. To derive co-occurrence probability (the link probability between two nodes), a local probabilistic graph model using Markov Random Fields (MRF) is proposed. To predict whether two nodes x and y will be linked, there are three steps: 
(1) Use topological information to identify the central neighborhood set of $x$ and $y$.
(2) Select itemsets that lie entirely within this set and use them as training data to train a local probabilistic model. Here, the training process is translated to a maximum entropy optimization problem.
(3) Estimate the co-occurrence probability features through inference over the local model.
Then Logistic Regression is used as a classifier to train the data which combines the above three types of features.
\begin{figure}[!htb]
\centering
\includegraphics[width=0.7\textwidth]{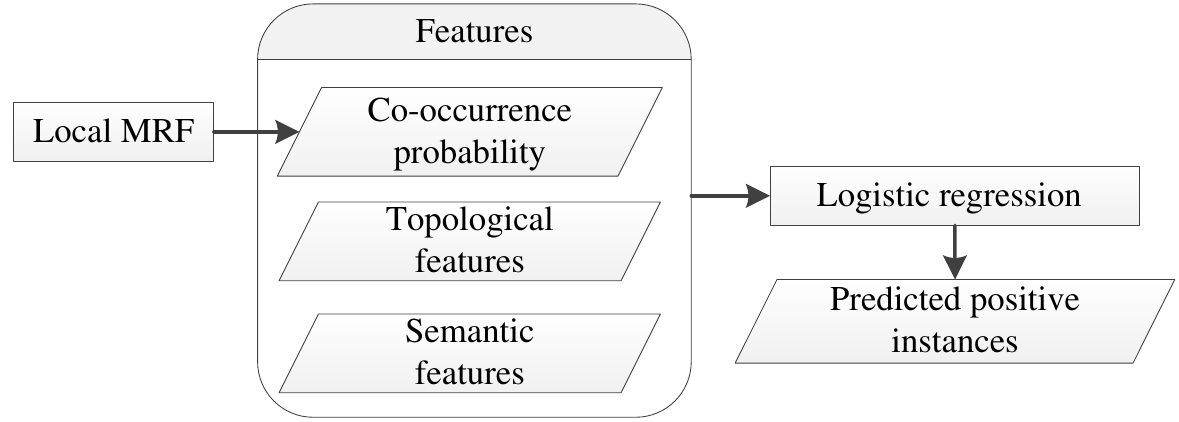}
\caption{Local probabilistic model.}
\label{fig11}
\end{figure}

Kashima et al. \cite{KA06} proposed a parameterized probabilistic model of network evolution to predict whether an edge between two nodes exists or not. In this model, the structure of a network changes probabilistically over time, namely, the edge label function  $\phi$ changes over time. An edge label $\phi(x,y)$ indicates the probability that an edge exist between x and y. $\phi^{(t)}$ denotes label function at time $t$. The model is assumed to be a Markov model in which $\phi^{(t+1)}$ depends only on $\phi^{(t)}$.  Assume that node $x$ has a strong influence on $y$ and there is a link between $x$ and $z$, so there will likely be a link between $y$ and $z$. In the model, there are two ways for $\phi^{(t)} (x,y)$ to assume a particular label. One way is that node $k$ has copied an edge label to node $x$ or node $y$. The other is that $\phi^{(t)} (x,y)=\phi^{(t+1)} (x,y)$, namely, nothing has happened. Then based on the assumption that the current network is a stationary state of the network evolution, an Expectation Maximization (EM)-type transductive learning approach is employed to estimate parameters of the model for predicting the existence of links.
Therefore, the basic idea behind this model is: if you have a friend who has a strong influence on you, your association will be highly affected by the friend's association.
This approach is based on the topological features of network structures and not on the node features. It differs from these earlier works in that the existence of tunable parameters within the network model naturally gives rise to a learning algorithm for link prediction, leading to improved accuracy of prediction. This model can be easily generalized to the scenario in which there are more than two edge labels.

Yang et al. pointed that the information contained in interest networks and friendship networks is highly correlated and mutually helpful. Based on homophily, a friendship-interest propagation framework is proposed for linking a user to interested services and connecting different users with common interests \cite{YLS11}. The framework devises a factor-based random walk model to explain friendship connections, and simultaneously it uses a coupled latent factor model to uncover interest interactions .

Backstrom and Leskovec \cite{BL11} proposed an algorithm for link prediction based on supervised random walks. Unlike traditional PageRank assumes the same transition probabilities of all links, supervised random walks learn a function to assign different transition probability for each link so that the random walk is more likely to visit target nodes than other nodes of the network. The function (edge strength) is computed based on the attributes of node $x$ and $y$, as well as the attributes of the link between $x$ and $y$. In this way, the approach can combine structural information of the network with the attributes of nodes and links for link prediction. As Figure 12 shows, to predict new edges of a given source node $s$, there are three steps:
(1) Using edge strength function to calculate the edge strengths of all edges.
(2) A random walk with restarts is run from $s$, and the stationary distribution of the random walk assigns each node a probability.
(3) A rank of nodes is generated by the order of this probability and the top ranked nodes are then predicted as destinations of future link links.
Compared to supervised machine learning methods, this approach does not require complex network features and domain related knowledge.

\begin{figure}[!htb]
\centering
\includegraphics[width=0.7\textwidth]{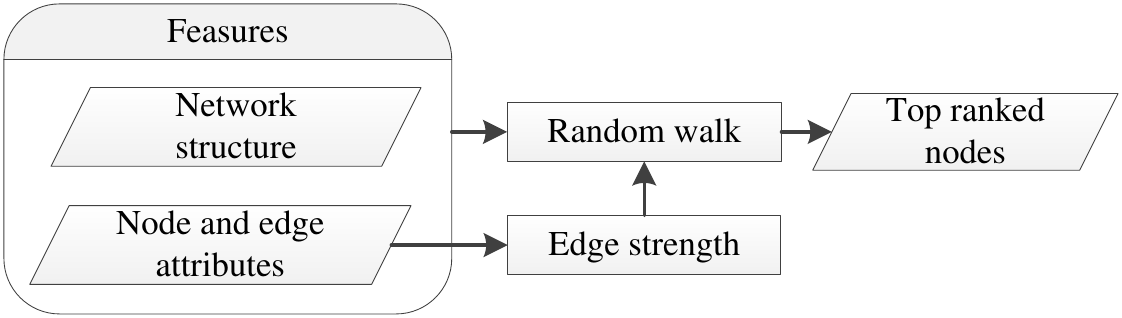}
\caption{ Link prediction based on supervised random walks.}
\label{fig12}
\end{figure}

The structural information is useful for link prediction in a hybrid network \cite{YHD11}. Link prediction can also be solved by mining graph evolution rules \cite{BBB10}. 
The real-world network is usually incompletely observed. To predict which of the possible unobserved links are actually present in the network, Marchette and Priebe apply a constrained random dot product graph to rank the potential edges according to the probability that they are in fact present, and then utilize covariates to improve the link prediction \cite{MP08}.
In link prediction, if the network itself is totally missing, namely, without the knowledge of an existing link structure, while some other information regarding the nodes is available such as interest group and tags, it is called the cold start link prediction problem. Leroy et al discussed this problem and proposed a two-phase solution based on the bootstrap probabilistic graph\cite{LCB10}. The first phase generates an implicit social network under the form of a probabilistic graph. The second phase applies probabilistic graphs-based measures to produce the final prediction. 

For the reason of privacy, not all social networks provide labeled data such as ``who like a restaurant''. Especially in anonymous social networks like Foursquare and Secret App, only the aggregative statistics information like ``how many people like this restaurant'' is available. 
To predict the opinion holder in such heterogeneous social network without labeled data, Kuo et al. generalized it to a link prediction with aggregative statistics problem and proposed an unsupervised probabilistic graphical model to solve it \cite{KYH13}. First, a factor graph model with three layers of random variables infers the existence of unseen-type links. Then three types of potential functions integrate diverse information into the factor graph model. A ranked-margin learning algorithm tunes the parameters using aggregative statistics. Finally, a two-stage inference algorithm updates potential functions and optimizes the results.  

Table 5 shows the comparison of some probabilistic graph models in link prediction. It can be seen that most methods use or modify existing probabilistic graph models such as random walk and factor graph model. Moreover, though most methods are suitable for general networks, some methods are devised for special networks such as hierarchical networks and incomplete networks. Like other kinds of learning-based methods, probabilistic graph models achieve better performance than basic topology-based metrics, especially improve the prediction accuracy.  Finally, since the probabilistic graph models exploit global network information, some methods are scalable to large scale networks.

\begin{table}[!htb]
\centering
\caption{Comparison of probabilistic graph models}
\footnotesize

\begin{tabular}{p{70pt}|p{105pt}|p{80pt}|p{150pt}}
\toprule
  Methods & Graph models & Network types & Characteristics\\
  \hline
Clauset et al.\cite{CCN08} & Hierarchical random graph, maximum likelihood, Monte Carlo sampling &Hierarchical networks& Accurately predict missing links; performs poorly for networks have no hierarchical structure\\
 \hline
Guimer\`aa and Sales-Pardo\cite{GS09} &Stochastic block model & Noisy networks & Outperforms at identifying both missing links and spurious links; high computation time\\
\hline
Chen and Zhang \cite{CZ14}& Marginalized denoising model  & General large networks & Models the dense and smooth affinity matrices; is scalable to large networks\\
\hline
Wang et al. \cite{WSP07} & Maximum entropy Markov random fields & Co-authorship networks& Co-occurrence probability feature is effective for link prediction, and combining with topology features and semantic features can improve the performance \\
\hline
Kashima and Abe \cite{KA06} & Parameterized probabilistic model, incremental learning & Dynamic networks& Achieves better performance than basic topology-based metrics\\
\hline
Yang et al. \cite{YLS11} & Friendship-interest propagation framework devises a factor-based random walk model & Interest networks, friendship networks & Bridges collaborative filtering in recommendation systems and random walk\\
\hline
Backstrom and Leskovec \cite{BL11} & Supervised random walks & General networks& Combines network structure with the attributes of nodes and links; requires no network feature generation\\
\hline
Marchette and Priebe \cite{MP08} & Constrained random dot product graph & Incomplete networks & Predicts the possible unobserved links that actually present in the network \\
\hline
Leroy et al. \cite{LCB10} & Bootstrap probabilistic graph & Networks without the initial status and with other information&Handle the cold start link prediction: predicting the structure of a social network when the network itself is totally missing while some other information regarding the nodes is available\\
\hline
Kuo et al. \cite{KYH13} & Factor graph model  & Networks with aggregative statistics of links &Link prediction with aggregative statistics problem\\
 \bottomrule
\end{tabular}
\end{table}

\subsubsection{Matrix Factorization}

Menon et al. \cite{ME11} treated link prediction as matrix completion problem and extend matrix factorization method to solve the link prediction problem. They factorize the graph $G\approx L(U\land U^\text{T})$ for $U\in \mathbb{R}^{n\times k}$, $\land \in \mathbb{R}^{k\times k}$ and link function $L(\cdot)$, where $n$ is the number of nodes and $k$ is the number of latent features. Each node $x$ has a corresponding latent vector $u_x\in \mathbb{R}^k$. Then the model's predicted score for the pair $(x,y)$ is $L(u_x^\text{T} \land u_y)$. This model combines latent features with explicit features for nodes and links in the graph via a bilinear regression model. The latent features can be also combined with the results of any other link prediction models. The model optimizes for AUC directly in order to overcome the imbalance problem, which refers to the phenomenon of positive links account for a very small percentage of all link instances but negative links account for most of link instances.

\subsection {Datasets and Tools}

Almost all link prediction works need to verify their methods on the collected datasets. 
The datasets are important for fairly reproducing and comparing different link prediction methods.
Constructing and collecting the datasets is a time-consuming and labor-intensive work.
However, not all datasets are public and available.
During surveying the link prediction works, we summarize some popular datasets used in link prediction, which are shown in Table 6. 
Especially, all the datasets are open, public, available and real-world. 
Online social network platforms (such as Twitter and Facebook) and public bibliography libraries (such as DBLP and Arxiv), are among the most popular sources of datasets. 
We notice that some datasets are well maintained by Stanford University. This would be an important reason that many researchers prefer to use these datasets.
However, we have to point out some disadvantages of current datasets for link prediction.
First, some datasets have noise, that must be cleaned before they are used.
For example, all bibliographic networks are faced with the author name disambiguation problem, which will cause a lot of noise and make the networks inconsistent to the real-world networks. 
Second, these datasets are not rich and diverse enough in term of the size and network types.
For some new and special link problems or applications, it may be unable to find existing datasets.
Finally, when same metrics are compared on different datasets, their performance ranks are usually not consistent or even various greatly. 
However, current datasets are not enough for analyzing the strength and weakness of a link prediction metric or method. This would lead to misusing the link prediction methods.
Therefore, it is necessary to build and maintain the benchmark datasets for link prediction problems.

\begin{table}[!t]
\centering
\caption{Popular open datasets used in link prediction}
\footnotesize

\begin{tabular}{p{40pt}|p{83pt}|p{40pt}|p{33pt}|l}
\toprule
  \multirow{2}{*}{Data source} & \multirow{2}{*}{Description} & \multicolumn{2}{c|}{Size} & \multirow{2}{*}{Sites of datasets}\\
  \cline{3-4} & & Nodes & Edges & \\
  \hline
  
  \multirow{3}{*}{DBLP} & Co-authorship network &  317,080 & 1,049,866& snap.stanford.edu/data/com-DBLP.html \\

  \cline{2-5}& Paper citation network & 324,339 & 812,740& arnetminer.org/citation  \\
  
  \cline{2-5} & Heterogeneous bibliographic network &   28,702 authors; 28,569 documents  &103,201 & www.cs.uiuc.edu/~hbdeng/data/kdd2011.htm\\
  \hline
  
  \multirow{2}{*}{Arxiv} &\multirow{2}{3.3cm}{High energy physics paper citation network}  &\multirow{2}{*}{34,546} &\multirow{2}{*}{421,578} &  snap.stanford.edu/data/cit-HepPh.html\\
  \cline{5-5} & & & & snap.stanford.edu/data/cit-HepTh.html\\
  \hline
  
  NIPS 1-17 & Co-authorship networks in NIPS& 2,865&4,733 & ai.stanford.edu/~gal/data.html\\
  \hline
  
  \multirow{2}{*}{Enron email} &   \multirow{2}{3.3cm}{Email communication network} &28,000 & 250,000&www.cs.cmu.edu/~enron/\\
    \cline{3-5} & &36,692 & 183,831&snap.stanford.edu/data/email-Enron.html\\
  \hline
  
  Patents citation &Citation network among US patents & 3,774,768 &16,518,948 &snap.stanford.edu/data/cit-Patents.html\\
  \hline
  
  \multirow{3}{*}{Facebook} &\multirow{3}{3.3cm}{Interactions between users on Facebook} & 4,039& 8,8234& snap.stanford.edu/data/egonets-Facebook.html \\
  \cline{3-5} & &60,290&1,545,686 & socialnetworks.mpi-sws.org/datasets.html\\
  \cline{3-5} & & 3,694& 13,692& delab.csd.auth.gr/~symeon/facebook.txt\\
  \hline
  
  \multirow{2}{*}{Twitter} & \multirow{2}{3.3cm} {Interactions between users on Twitter} &81,306    &1,768,149 & snap.stanford.edu/data/egonets-Twitter.html \\
  \cline{3-5} & & 124,501& 22,169,689& lsir.epfl.ch/aberer/\\
  \hline
  
  Foursquare & Location-based social network & 269,279& 1,101,504&  www.csie.ntu.edu.tw/~d97944007/aggregative/ \\
  \hline   
  
  MovieLens &Movie rating network & 72,000 users; 10,000 movies & 10,000,000 ratings &grouplens.org/datasets/movielens/ \\
  \hline
  
  Book-Crossing   &Book ratings network &278,858 users; 271,379 books  &1,149,780 ratings & www2.informatik.uni-freiburg.de/~cziegler/BX/\\
  \hline
  
  Wikipedia & Vote network &7,115 &103,689 & snap.stanford.edu/data/wiki-Vote.html  \\
  \hline
  
  Epinions & Trust social network & 131,828 & 841,372 & snap.stanford.edu/data/soc-sign-epinions.html\\
  \hline
  
  Slashdot & Signed network &82,140 &549,202 &snap.stanford.edu/data/soc-sign-Slashdot090221.html\\
  \hline
  
  Plurk & Micro-blog social network & 543,329&3,660,507& www.csie.ntu.edu.tw/~d97944007/diffusion/\\

 \bottomrule
\end{tabular}
\end{table}

Although there are many link prediction metrics and methods proposed, only very few works open their source codes. People have to re-implement some complicate methods, and that is a time-consuming process.  Only few public tools try to integrate these metrics and methods. It is very important for selecting the appropriate metrics or methods for a link prediction task. 
For example, LPmade, which is a cross-platform software solution, provides multi-core link prediction and related tasks and analysis \cite{LC11}. First, it is a scalable library which implements most commonly used unsupervised link prediction metrics, especially, all implementations have high performances. Second, it supports automatic link prediction processes including prediction, evaluation, and network analysis.
If we have the benchmark datasets for link prediction, we could implement an open link prediction API with standard datasets loading, formatted results and automatic evaluation. 
It would also greatly reduce the evaluation work.

\section {Link Prediction Problems}
There are many works focused on solving special link prediction problems, which can be divided into six categories: temporal link prediction, active/unactive link prediction, link prediction in bipartite networks, link prediction in heterogeneous networks, unfollow or disappearing link prediction, and link prediction scalability.

\subsection{Temporal Link Prediction}

In recent years, the research on link prediction has evolved over various aspects. One is to consider the time in the model, which can be named as temporal link prediction \cite{DKA11,OHS05} . A social network with time can be organized as a third-order tensor, or multi-dimensional array. $A$ tensor $\mathcal{Z}$ of size $M\times N \times T$ can be defined as
\begin{equation}
\mathcal{Z}(i,j,t)=
\begin{cases}
1 & \textrm{if vertex $i$ links to vertex $j$ at time $t$} \\
0 & \textrm{otherwise}
\end{cases}
\end{equation}
It can answer specific questions such as ``Who is most likely to publish at ICDM next year''. Given social network for times 1 through $T$, it needs to predict the links at time $T+1$.

Dunlavy et al. proposed a novel method for temporal link prediction by combining matrix-based and tensor-based techniques \cite{DKA11}, and this method is showed in Figure 13. For the matrix-based part, it first collapses the network data into a single matrix. The collapsed tensor (CT) is produced by collapsing the network into a single matrix through summing all the entries across time. The improved collapsed tensor considers the weight when sum the slices in the time, and it is called the collapsed weighted tensor (CWT). CWT can give greater weight to more recent links. Then a matrix of scores for link prediction can be calculated by truncated SVD, and an extended Katz metric to the case of bipartite graphs and its relationship to the matrix SVD is also derived. Using the truncated SVD, this method is salable for calculating a truncated Katz score. However, there is no matrix-based method fully leverages and exposes the temporal information.
 
Tensor factorizations are higher-order extensions of matrix factorizations that capture patterns in multi-way datasets and have proved to be successful in diverse disciplines. 
Based on the classical tensor model CP, heuristic-based and forecasting-based prediction methods are proposed by using the temporal information extracted by CP. For the heuristic-based method, it defines the similarity score for vertex $x$ and $y$ using a K-component CP model. The forecasting-based prediction method provides a more sophisticated use of time, and it uses the temporal profiles computed by CP as a basic for predicting the scores in future. It is an automatic method which only requires the data and the expected period. The advantage of this tensor-based method is that it can better capture and exploit temporal patterns. The drawback of the tensor-based approach is its higher computational cost.

\begin{figure}[!t]
\centering
\includegraphics[width=0.9\textwidth]{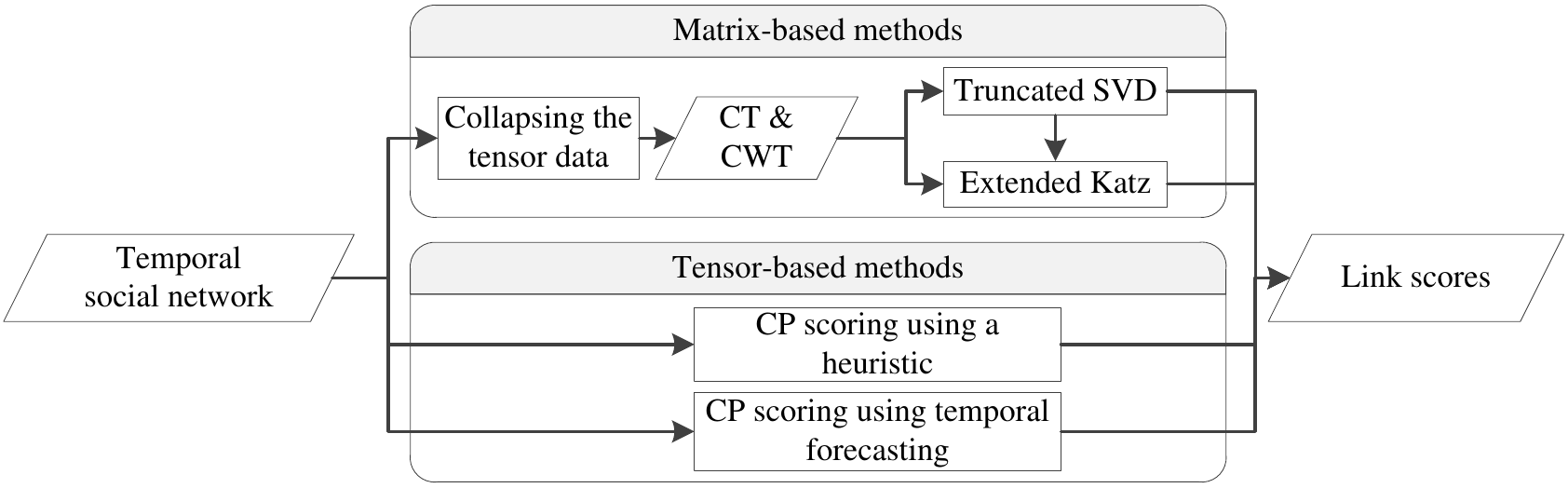}
\caption{The matrix and tensor-based temporal link prediction.}
\label{fig13}
\end{figure}

Gao et al. proposed a novel unified model based on latent matrix factorization method and graph regularization technique which integrates content and structure information to capture the time evolving patterns of links in the networks  \cite{GDG11A}. This method makes use of an efficient alternating iterative algorithm for learning the latent factors of nodes in the network, and it provides the possibility for handling large scale networks.

Tylenda et al.  developed graph-based link prediction techniques that incorporates the temporal information contained in evolving social networks \cite{TAB09}. They extended the predict model to include time awareness, and show how to incorporate edge weights which derived from temporal features into the state-of-the-art link prediction methods. 
Huang et al. investigate hybrid link prediction methods that combine the power of the time-series model in predicting repeated link occurrences with the ability of static graph link prediction methods to identify new link occurrences \cite{HJ09}. 
Soares and Prud\^encio et al. built time series for each pair of non-connected nodes by calculating their similarity scores at different past times  \cite{SP12}. Subsequently, they designed a forecasting model based on time series to obtain final link prediction scores of the pairs. However, according to above works, only links in identical time frames are considered in temporal link prediction. Oyama et al. proposed the cross-temporal link prediction method, in which the links among nodes in different time frames are inferred \cite{OHK11}, 
and a dimension reduction approach makes high dimensional data to be mapped to a low-dimensional latent feature space. 

Munasinghe and Ichise focused on the temporal behavior of the link strength, particularly the relationship between the time stamps of interactions or links and the temporal behavior of link strength and how link strength affects future link evolution\cite{MI12A}. Most previous studies have neither sufficiently discussed the impact of time stamps of the interactions nor time stamps of the links on link evolution.
It believes that the strength of a link varies over time. 
The nodes that do not interact with each other for a long time will cause the links become weaker. 
So higher scores are assigned to node pairs which have interacted with their common neighbors recently. 
In other words, if the difference between the time stamps of the most recent interactions of common neighbors having the node pair is small, then this difference has a greater effect on future links. Combining the above considerations, a new feature called \textit{time score} (TS) is introduced, to take into account the time awareness for link prediction. Time score for the node pair $x$ and $y$ that has $n$ common neighbors is defined as follows:
\begin{equation}
\text{TS}(x,y)=\sum_{C_i\in\mit\Gamma(x)\cap\mit\Gamma(y)}\frac{H_m^i\beta^{k_i}}{|t_1^i-t_2^i|+1}
\end{equation}
where $c_i$ is the common neighbors of $x$ and $y$, $t_1^i$ is the most recent time stamp of the interactions between $x$ and $c_i$, 
$t_2^i$ is the most recent time stamp of the interactions between $y$ and $c_i$, $\beta$ is a damping factor $(0<\beta<1)$, $k_i$ is the difference between current time $t_c$ and the most recent time stamp form $t_1^i$ and $t_2^i$, and $k_i$ is defined as: $k_i=t_c-max(t_1^i-t_2^i)$.
$H_m^i$ is the harmonic mean of the cooccurence frequencies of $x$ and $y$ with the common neighbor $c_i$.
The effectiveness can be affected for time score with different parameter settings for different network datasets, and the time score is sensitive to different networks and different time measures.
It can be seen that time score incorporates the impact of the time stamps of the interactions and the gap between the current time and the time stamps. 
Time score feature can be used in the supervised machine learning methods for predicting links. 

Juszczyszyn et al. defined the Triad Transition Matrix (TTM) containing the probabilities of transitions between 64 triads found in the network, and proposed its application for link prediction with an algorithm (called TTM-predictor) which shows good performance, especially for sparse networks analyzed in short time scales \cite{JMB11}. Since the TTM is based on the temporal network to calculate the transition probabilities of triads, it can be seen as a kind of temporal link prediction method. 

Soares and Prud\^encio proposed a link prediction measure based on temporal events \cite{SP13}. The event-based score is updated along time by rewarding the temporal events observed between the pair of nodes under analysis and their neighborhood. The dynamics of links as the network evolves is used to update representative scores to pairs of nodes, rewarding pairs that formed or preserved a link and penalizing the ones that are no longer connected.

Richard et al. investigated the links of the graph and its topological features that have been evolving over time may also be useful to predict future links \cite{RBE10}. Their work lies in the observation that a few graph features that can capture the dynamics of the graph evolution and provide information for predicting future links. The main idea is to learn over time the evolution of well-chosen local features (at the level of the vertices) of the graph, and then uses the predicted value of these features on the next time period to discover the missing links.

Jahanbakhsh et al. exploited time-spatial properties of contact graphs as well as the popularity and social information of mobile nodes to propose a method for reconstructing the missing parts of contact graphs in mobile social networks\cite{JKS12}.

\subsection{Link Prediction in Heterogeneous Networks}

Most of the existing link prediction works focus on homogeneous networks, in which only one type of nodes or links exists. However, many social networks contain different link types and different kinds of nodes, which may have different typologies or link formation mechanisms and influence each other. Moreover, most of the networks in real world are heterogeneous and complete attribute values of nodes are often difficult to obtain. Therefore, link prediction on such heterogeneous social networks is also a non-trivial task.  

Heterogeneous relationships such as friendship, family, and colleague are often modeled as indistinct in social networks. Several new multi-relational link prediction (MRLP) methods for heterogeneous information networks are proposed to overcome shortcomings of traditional prediction methods \cite{LLC13,GDG12,GDG11B}. The key component of the MRLP is an appropriate weighting scheme for different edge type combinations. 
The weights are determined by counting the occurrence of each unique triad census with three nodes.
The triad census also provides the probability of each structure, which further translates to the probability that a partial triad is closed by respective edge type. 
Specifically, for three nodes $(s,n,t)$ and an edge type $x$, it first counts all triads with the same pattern as $(s,n,t)$, then counts all triads with the same pattern with $x$ added between $s$ and $t$. 
$P(x\in edge\_type(s,t)|pattern(s,n,t))$ is determined by dividing the first count by the second count. This probability assumes that the observed pattern is correct except for the potential absence of type $x$, which simplifies the calculation. 
Rather there is no unsupervised method that consistently performs well and MRLP outperforms AA metric in most cases. 
The supervised framework is more consistent in its performance, generally achieving higher scores than unsupervised methods.

Sun et al. considered a heterogeneous bibliographic network \cite{SBG11,SHA12}, which is different from the traditional co-author network and contains multiple types of objects, such as authors, venues, topics and papers, as well as multiple types of links denoting different relations among these objects, such as ``write'' and ``written by'' relations between authors and paper, ``cite'' and ``cited by'' relations between papers, and so on. To predict the relationship building time between two objects, namely, whether or when a relationship between two objects will be built, target relation and topological features are encoded in a meta-path, then a generalized linear model based supervised framework is used to model the relationship building time. Sun and Yang et al. also proposed a new topological feature called multi-relational influence propagation to capture the correlation between different types of links and then proposed temporal features in heterogeneous networks to achieve better link prediction accuracy \cite{YCS12}.

Davis et al. proposed a novel MRLP method for heterogeneous information networks to predict the location and type of new edge \cite{DLC11}. The key idea of this method is an appropriate weighting scheme for different edge type combinations. The weights are determined by counting the occurrence of each unique triad census in networks. The triad census provides the probability of each structure. This MRLP method can be seen as a weighted extension of the neighborhood methods.

Since that people in different social networks would interact thus predicting links across heterogeneous social networks is also an interesting problem. 
Dong et al. investigated the link formation over different social networks and  find some interesting general social patterns in triad relationships according to social theories including degree distribution, social balance and microscopic mechanism \cite{DTW12}. Then they defined social pattern-based features and proposed a transfer-based ranking factor graph model for the link prediction across heterogeneous networks.  

Link prediction in heterogeneous networks is still an open problem, and a lot of issues need to be discussed.
Str\"oele et al. proposed a composite metric based on three basic metrics: node degree, common neighbors and Katz metric, to predict new relationships in scientific social networks \cite{SZS13}. Other researchers have also discussed the scalability of link prediction in heterogeneous social networks \cite{RBG11}. Wang et al. found that human mobility could indeed serve as a good predictor for the formation of new links, yielding comparable predictive power to traditional network-based measures. By combining both mobility and network measures, they showed that the prediction accuracy can be significantly improved in supervised learning \cite{WPS11}.

\subsection{Link Prediction with Active and Unactive Links}

Munasinghe and Ichise gave an assumption that if a node pair interacts recently, then the link between them becomes active \cite{MI12B}. The time stamp of the last interaction is a vital information in deciding the activeness of a link. Hence, the most recent time stamps of the interactions between nodes is used in link prediction computations. T\_Flow based on the PropFlow metric is proposed. It considers link weight as well as link activeness to compute transition probabilities. T\_Flow imports a decaying function to describe the decay of information. For two adjacent nodes $x$ and $y$, the decaying function is defined as $d(x,y)=(1-\alpha)^{|t_x-t_y |}$, $0< \alpha <1$. Therefore, the T\_Flow from node $x$ to $y$ via direct link can be calculated as follows:
\begin{equation}
\text{T\_Flow}(x,y)=\text{T\_Flow}(a,x)\cdot\frac{w_{xy}}{\sum_{k\in\mit\Gamma(x)}w_{xk}}\cdot d(x,y)
\end{equation}
If nodes $x$ and $y$ are indirectly linked, it computes the T\_Flow through all the shortest paths from node $x$ to $y$ recursively and takes the summation. The total flow between two nodes is regarded as the T\_Flow for the node pair. T\_Flow outperforms the PropFlow in social networks with active and unactive links.

Chen et al. observed an interesting phenomenon in FOAF social networks: ``younger links seem to be more influential in future link prediction''\cite{CGZ12B}. It implies that recent links are more important than older ones in link prediction. 
To verify this observation, a new relation strength similarity (RSS) metric is applied on a co-authorship network to study the power of recency. RSS metric is one of the few similarity measures designed for weighted networks and easily models FOAF networks. By assigning different weights to the links according to authors' coauthoring history, it shows that recency is helpful in predicting new links.
The new relation strength considering both the number of coauthored papers and the recency factor between $v_i$ and $v_j$ is defined as
\begin{equation}
R(v_i,v_j)=\frac{n_{i,j}(t_{now})}{\sum_{\forall k\in\mit\Gamma(v_i)}n_{i,k}(t_{now})}
\end{equation}
where $n_{i,j} (t_{now})$ is the edge weight at time $t_{now}$. Then the general relation strength from $v_i$ and $v_j$ can be calculated by RSS metric.
Chen et al. also applied a supervised learning approach to study link age as a factor for link prediction \cite{CMG13}. Unlike those that suggest a relatively ad-hoc aging model, here they apply logistic regression to quantify the relative importance of old links and young links. The experiments on several real world datasets show that younger links are more informative than older ones in predicting the formation of new links. Since older links become less useful, it might be appropriate to remove them when studying network evolution.

\subsection {Link Prediction in Bipartite Networks}

Many social networks are bipartite networks such as the user-product networks in e-commerce area.
However, the link prediction problem is usually defined on unipartite graphs, where common link prediction methods make several assumptions: (1) triangle closing: new edges tend to form triangles; (2) clustering: nodes tend to form well-connected clusters in the graph. But in bipartite graphs these assumptions are not true, since triangle and larger cliques cannot appear. While a unipartite link prediction method applies to bipartite graphs, it will not perform well. Fortunately, there are some researchers pay attention to this issue.

Kunegis et al. found that for the simple local link prediction methods, only the preferential attachment model can be used in bipartite networks \cite{KLA10}. Algebraic link prediction methods can be used instead, by restricting spectral transformations to odd functions, leading to the matrix hyperbolic sine as a link prediction function, and an odd variant of the von Neumann kernel.

Some researchers extend the classical link prediction methods such as common neighbors, Jaccard coefficient, Adamic Adar, and Preferential Attachment metrics to bipartite networks\cite{XDL12,CK12}. The key idea is using neighbors's neighbors to replace the direct neighbors. Xia et al. studied the link prediction in bipartite social networks, and then proposed two measures of structural holes for link prediction in bipartite networks: one is absent links, another is minimum description length \cite{XDL12}. Allia et al. transformed bipartite graphs into classical graphs by projection, then introduce internal links in bipartite graphs for link prediction \cite{AML11}. 
Liu and Deng discussed the link prediction in user-object network, which is a more abstract bipartite network. Based on the resource allocation method, they proposed a time-weighted network to model the evolution of the user-object network \cite{LD09}. It shows that both time attenuation and diversion delay play key roles in link prediction in a user-object network.

\subsection{Link Prediction for Unfollow or Disappearing Links}

The formation and dissolution of link are two fundamental processes of link change and evolution in dynamic networks. Links in social networks could be appeared or disappeared. For example, user A in Twitter breaks the relationship with another user B. In social networks, we call this behavior ``unfollow''.
To the best of our knowledge, numerous efforts have been made in studying link formation for predicting new links in future, but only few attentions are paid to link dissolution, namely, predicting the disappearing link in future.

Kwak et al. analyzed the unfollow behavior in Twitter, including the characteristics of the unfollow behavior and the reasons why people unfollow each other. They found Twitter users frequently unfollow, and discover some major unfollow factors including the reciprocity of the relationships, the duration of a relationship, the followees's informativeness and the overlap of the relationships\cite{KCM11}. They also took other factors into consideration \cite{KML12}, including individual, dyadic and triadic properties between ego and alter of the link, and use these factors to build a logistic regression model. From the fitted model, they found some structural and actional factors can significantly explain the unfollow behavior. Later, the same research group used actor-oriented modeling (SIENA) to examine the impacts of reciprocity, status, embeddedness, homophily, and imformativeness on the unfollow behavior \cite{XHK13}. The results show mutual following relationship and common followees reduce the likelihood of the unfollow behavior. 

Kivran-Swaine et al. explored network structure alone can significantly influence link dissolution in Twitter. They use social theory, such as link strength, embeddedness, power and status, to study the unfollow behavior\cite{KGN11}. Quercia et al. considered factors that study in sociology (age, gender, and personality traits) to study the unfollow behavior in Facebook. They found that the link is more likely to break if it is not embedded in the network, if it is between two people whose ages differ, and if one of the two is neurotic or introvert \cite{QBC12}.

\subsection {Link Prediction Scalability}
The scalability and effectiveness are both important for massive real world social networks.  
Sarkar et al. proposed a nonparametric link prediction for dynamic networks \cite{SCJ12} in which their model can accommodate regions with very different evolution profiles, otherwise impossible by the link prediction metric or heuristic. It also enables learning based on both topological as well as other externally available features. They also adapted the locality sensitive hashing algorithm to solve the scalability for link prediction in large networks and long time sequences.
Song et al. develop two novel methods to efficiently and accurately approximate a large family of proximity measures, which is a challenge for massive scale and dynamic online social networks. Then the proposed proximity estimation is used for link prediction, and obtains high link prediction accuracy by combining multiple proximity measures \cite{SWV09}. A novel incremental update algorithm is proposed to enable near real-time link prediction in highly dynamic social networks. 

To handle the link prediction in massive evolving networks with sparse connectivities and nonlinear transitional patterns, 
Li et al. proposed a deep learning framework called conditional temporal restricted Boltzmann machine \cite{LDL14}, which predicts links based on individual transition variance and influence introduced by local neighbors. 
It is robust to noise and has the exponential capability to capture nonlinear variance. 
Besides the computational benefits, this method devises two types of directed connections to the hidden variables: temporal connections and neighbor connections, which fit two well known assumptions: each node has a unique transitional pattern and a node's behavior is influenced by its local neighbors. 

\section {Link Prediction Applications}
In social networks, link prediction can be used for various applications; here we will address some typical applications, such as recommendation in social networks, network completion, and social ties prediction.

\subsection{Recommendation in Social Network}

Recommending partners, friends, followees and followers is a typical application for link prediction.
Wu et al. proposed a novel interactive learning framework to formulate the problem of recommending patent partners into a factor graph model \cite{WST13}. 
Wu and Dong et al. also developed a transfer-based factor graph model that combines them with network structure information for link recommendation across heterogeneous social networks \cite{DTW12}.  
Armentano and Godoy proposed a followee recommender system based on both the analysis of the content of microblogs to detect users's interests and in the exploration of the topology of the network to find candidate users for recommendation \cite{AGA13}. They found that user-generated content available in the network is a rich source of information for profiling users and finding like-minded people.
Sadilek et al. infered social ties by considering patterns in friendship formation, the content of people's messages, and user location \cite{SKB12}. They first employed text similarity between users's tweets, co-location and neighbor-based graph structure as features, then use a Markov random field model to learn and inference friendship prediction.
Huang et al. viewed user-item interactions in recommender system as graphs and employ link prediction approaches for making collaborative filtering recommendations \cite{HLC05}. They adopted some common similarity-based link prediction methods and find these methods can achieve significantly better performance than standard collaborative filtering methods.
Rowe et al. predicted follower edges within a directed social network by exploiting concept graphs and different behaviors that users exhibit \cite{RSA12}. Their method significantly outperforms a random baseline and models that rely solely on network topology information. 
Link prediction can also improve the quality of behavioral recommender \cite{EBB11}.

\subsection{Reciprocal Relationship Prediction}

In social networks, a two-way (also called reciprocal) relationship, usually developed from a one-way (parasocial) relationship, represents a more trustful relationship between people. Understanding the formation of two-way relationships can provide us insights into the micro-level dynamics of the social network, such as the underlying community structure and users's influence on each other.
Hopcroft et al. studied the extent to which the formation of a two-way relationship can be predicted in a dynamic social network \cite{HLT11}. 
A semi-supervised learning framework is proposed to formulate the problem of reciprocal relationship prediction into a Triad Factor Graph model, which incorporates social theories. A large Twitter network is used to verify this framework. 
The method can accurately infer 90\% of reciprocal relationships in a dynamic network. 
In addition, it provides strong evidence of the existence of the structural balance among reciprocal relationships. More importantly, the results have potential implications such as how social structures can be inferred from individuals' behaviors.

\subsection{Network Completion Problem}

Usually, the collected social network data is incomplete with nodes and edges missing. Since that only a part of the network can be observed or collected, it needs to infer the unobserved part of the network. This is the network completion problem, wherein, given a network with missing nodes and edges, one has to  complete the missing part.
Tang et al. used the Expectation Maximization (EM) framework to model the social network completion problem, where the observed part of the network is used to fit a model of network structure, and then estimates the missing part of the network using the model, re-estimate the parameters and so on. They combined the EM with the Kronecker graphs model and devised a scalable Metropolized Gibbs sampling for the estimation of the model parameters and the inference about missing nodes and edges of the network \cite{TWS12}.
This method can effectively recover the network even half of the nodes in the network are missing. Especially, this method can  easily scale to large scale networks.

\subsection{Finding Experts and Collaborations in Academic Social Network}

Academic social networks contain massive amounts of experts in various disciplines and it is difficult for the individual researcher to decide which experts will match his own expertise best. 
Pavlov and Ichise propose a method for building link predictors in academic social networks, where nodes can represent researchers and links represent collaborations  \cite{PI07}.
It uses a supervised learning method for building link predictors from structural attributes of the underlying network. In a network of researchers, where a link represents a collaboration, such predictors could be useful in suggesting unrealized collaborations and thus help in building and maintaining strong research teams.
Then an improved method extracts structural attributes from the graph of past collaborations along with semantic and event-based features, and uses them to train a set of predictors using supervised learning algorithms \cite{WI08}. These predictors can then be used to predict future links between existing nodes in the graph. 

Interdisciplinary collaborations are valuable in human society. Establish cross-domain collaborations is difficult for some reasons: (1) cross-domain collaborations are rare; (2) cross-domain collaborators often have different expertise and interest; (3) cross-domain collaboration focus on a sub-topics.
Therefore, cross-domain collaborations have different patterns compared to traditional collaborations in the same domain.
Through analyzing the cross-domain collaboration networks, Tang et al. proposed the Cross-domain Topic Learning (CTL) model \cite{TWS12} to solve above challenges: (1) for handling sparse connections, CTL consolidates the existing cross-domain collaborations through topic layers instead of at author layers, which alleviates the sparseness issue; (2) for handling complementary expertise, CTL models topic distributions from source and target domains separately, as well as the correlation across domains; (3) for handling topic skewness, CTL only models relevant topics to the cross-domain collaboration.

Link prediction between co-authors is a frequently studied problem. Sun et al. studied the problem of co-author relationship prediction in the heterogeneous bibliographic network, and a new methodology called PathPredict based on meta path relationship prediction model was proposed to solve this problem \cite{SBG11}. First, meta path-based topological features are systematically extracted from the network. Then, a supervised model is used to learn the best weights associated with different topological features in deciding the co-author relationships.
On the other hand, the focus of traditional link prediction tasks is on the fact about whether a link will happen in the future, e.g., whether two people will become friends. However, in many applications, it may be more interesting to predict when the link will be built. Sun et al. proposed a new method to handle the problems such as ``what is the probability that two authors will co-write a paper within 5 years?'' \cite{SHA12}.

\subsection{Social Tie Prediction}

When a social network dynamically changes, the social ties would change over time. The social-tie strengths are different one another even though they are in the same group. Zhang and Dantu investigated the evolution of person-to-person social relationships, quantify and predict social tie strengths based on call-detail records of mobile phones \cite{ZD10}. They propose an affinity model for quantifying social-tie strengths in which a reciprocity index is integrated to measure the level of reciprocity between users and their communication partners. Some works focus on the relationship strength in social networks, such as predicting tie strength with social media\cite{GK09} and modeling relationship strength in online social networks \cite{XNR10}. Though these works do not directly relate to the link prediction, however, their results and conclusions can be extended to new link prediction methods.

\section{Active Research Groups}

Numerous efforts have been made by researchers from different institutions. This section will briefly address some works of research groups and their primary contributions on the link prediction problem.

\begin{itemize}
\item \textbf{Stanford University}
\end{itemize}

Leskovec et al. developed a concept of supervised random walks. It combines the network structure with the features of nodes and edges of the network into a unified link prediction algorithm \cite{BL11}. Then they develop a method based on it. The method learns to segregate a PageRank-like random walk on the network in a supervised way, so that it is more likely to visit nodes to which new links will be create in the future. Relationship can be either positive (friendship) or negative (opposition) in social networks, a model incorporating theories of balance and status from social psychology is used to predict the signs of relationships in social networks  \cite{LHK10}. To combine the analysis of signed networks with machine learning techniques, two categories of features are used. One is based on the degree of nodes and another is based on the principle from social psychology. Also, they investigate the network completion problem where nodes and edges in networks are both missing. They also develop KronEM, an EM approach combined with the Kronecker graphs model, to estimate the missing part of the network \cite{KL11}. 
Moreover, Leskovec et al. collected and constructed a lot of social network datasets which are public for other researchers.  These datasets have been used in many link prediction works. 

\begin{itemize}
\item \textbf{Tsinghua University}
\end{itemize}

Hopcroft and Tang's team  studies the novel problem of reciprocal relationship prediction to predict who will follow you back in directed social networks \cite{HLT11}. They proposed a Triad Factor Graph (TriFG) model, which incorporated social theories (such as structural balance and homophily) over triads into the semi-supervised machine learning model. Tang's team also formulated prediction problem to predict the existence and the type of links between a pair of nodes. They proposed a partially-labeled pairwise factor graph model (PLP-FGM) \cite{TZT11} and two active learning strategies (Influence-Maximization Selection and Belief-Maximization Selection) to capture the inter-relationship influence \cite{ZTT12}. They also extended the above model for the problem of inferring social ties across heterogeneous networks \cite{TLK12}. The model incorporates social theories into a semi-supervised learning framework, which can be used to transfer supervised information from a source network to help infer social ties in a target network. For the inventor social network where the link between inventors is the co-invention relationships. They also incorporate users's interactions into a factor graph model for recommending patent partners \cite{WST13}. This method shows good prediction accuracy and efficiency, so it could be beneficial for existing recommendation models based on users's feedback.

\begin{itemize}
\item \textbf{IBM Research}
\end{itemize}

Based on the topological features of network structure, Kashima et al. presented a parameterized probabilistic model of network evolution for supervised link prediction \cite{KA06}. In this model, the existence of links are modeled by a ``copy-and-paste'' mechanism. Then a link prediction algorithm is proposed based on the assumption that the network structure is in a stationary state of the network. Later, Kashima et al. develop a semi-supervised learning model for link prediction problem by applying label propagation to link prediction \cite{KKY09}. Label propagation is originally intended for use in node classification. They apply the idea of label propagation to pairs of nodes with multiple link types and predict the relationship among nodes. Besides link strengths, the model also can handle various types of links. And they utilized the matrix factorization techniques to solve the problem of computational time and space constraints here in \cite{RK10}. The new method can be regarded as node information-based prediction that utilizes fast and scalable techniques in topological-based prediction. They also extended link prediction to a more general form called cross-temporal link prediction in which the links among nodes in different time frames are inferred \cite{OHK11}. They adopted a dimension reduction approach where data objects in different time frames are mapped into a low-dimensional latent feature space.

\begin{itemize}
\item \textbf{University of Notre Dame}
\end{itemize}

Lichtenwalter et al. presented a flow-based predicting algorithm PropFlow \cite{LLC13}, which is proportional to the probability based on a restricted random walk. It is similar to Rooted PageRank, but it is more localized and does not need walk restarts or convergence. Due to heterogeneous links and complicated dependency structures in complex networks, they introduce a probabilistically weighted extension of Adamic-Adar coefficient for heterogeneous networks to model the influences between heterogeneous links and distinguish the formation mechanisms of each link type\cite{DLC11}. Also, they point out that the current methods of link prediction evaluation are inadequate and may lead to wrong conclusion about practical performance, so they propose a fairer framework and provide some guidelines for link prediction evaluation in \cite{LC12a}. Due to the extreme imbalance of link prediction, they argue for the use of threshold curves, such as precision-recall and AUC, rather than fixed-threshold metrics.
Lichtenwalter and Chawla also proposed the concept of vertex collocation profile (VCP), which is used in supervised models to discriminate potential new links \cite{LC12b, LC14}. They also developed LPmade, which is a complete cross-platform software solution for multi-core link prediction\cite{LC11}.

\begin{itemize}
\item \textbf{UESTC}
\end{itemize}

Researchers in University of Electronic Science and Technology of China (UESTC) present some link prediction metrics based on topological information for the abstract complex network.

RA metric \cite{ZLZ09}, which is motivated by the resource allocation process taking place on networks, is proposed and shown to have a similar form with AA metric. But it weights common neighbors differently. LP metric \cite{LJZ09}, which is based on local paths, also exploits information of the next nearest neighbors and can enhance the prediction accuracy compared to CN metric. Especially, LP metric is efficient in the hug network. To handle the sparsity and huge size of the network, they propose two similarity metrics for link prediction based on local random walk \cite{LL10}: LRW (Local Random Walk) and SRW (Superposed Random Walk), which can give competitively good prediction and low computation complexity. 

They also proposed a probabilistic model called local na\"ive Bayes (LNB) based on the Bayes theory \cite{LZL11}. Different to traditional methods in which each common neighbor contributes equally to the link likelihood, LNB considers that different common neighbors may play different roles in link prediction. The characteristic of the model is that two node pairs with same number of common neighbors could have different connection likelihoods.

They developed the weighted version of some similarity measures and find the weighted version perform worse. The experimental study shows that the weak ties play a significant role in the link prediction problem, especially for remarkably enhance the predicting accuracy \cite{LZ09}. However, most of their works only consider the link prediction problem in static networks.

Usually, in link prediction, the dataset is randomly divided into two parts: the training set and the probe set. It seems a fair method without statistical bias. They pointed out that such a straightforward and standard method may lead to bias, since missing links (the existed yet unknown links) are more likely to be links connecting low-degree nodes. Then they divided the dataset into two parts and make the links in the probe set less popular than the links in the training set. The experimental results show that the Leicht-Holme-Newman (LHN) \cite{LHN06} metric performs the best although it was known to be one of the worst metric if the probe set is a random sampling of all links. They further proposed a parameter-dependent metric to improve the prediction accuracy considerably\cite{ZLZ12}, namely, a similarity index with a free parameter $\lambda$, which depends on the average link popularity of the probe set. Through tuning $\lambda$, this metric can degenerate to CN metric, SI metric and the LHN metric. 

\begin{itemize}
\item \textbf{UIUC}
\end{itemize}

Yin et al. in University of Illinois at Urbana-Champaign (UIUC) proposed a random walk framework named LINKREC on an augmented social graph using both user attributes and structural information to predict links in social networks \cite{YGW10}. Besides person nodes, the augmented social graph contains additional nodes called attribute nodes. For every attribute of a person, a corresponding link between the person node and the attribute node is created. Then, they studied the problem of link prediction in heterogeneous network where contains multiple types of objects and links \cite{SBG11}. In heterogeneous networks, different paths between the same nodes may represent different relations and have different meanings. So they proposed a meta path-based relationship prediction model called PathPredict to solve the problem. 
 In \cite{SHA12}, they extended the above problem from ``whether it will happen'' to ``when it will happen'' and proposed a generalized linear model based supervised framework to solve the relationship building time prediction problem. The building time is treated as independent variables and their expectation is modeled as a function of topological features according to several reasonable distributions.

\begin{itemize}
\item \textbf{NII}
\end{itemize}

The research group in National Institute of Informatics (NII) of Japan tried to find experts by link prediction in co-authorship network using supervised learning algorithm. In \cite{PI07}, they used structural attributes extracted from the graph of past collaboration to train a set of predictors. Considering researchers's semantic descriptions might be helpful, besides the structure of the graph, they also add a semantic and event based features to improve the accuracy \cite{WI08}. The feature named keywords matching counting (KMC) represents the number of words in common between titles of their previous papers. And later, they made use of the fact that researchers tend to work in dense communities, and use community alignment information to further improve the accuracy of link prediction \cite{SI11}. Focus on the temporal behavior of link strength, they proposed a new time-aware feature called time score for link prediction. The feature incorporates the effectiveness of common neighbors and their temporality by assigning higher score to node pairs which interact with their common neighbors within more recent time \cite{MI12A}. They also extend this study, which is limited to nodes pairs having common neighbors, to any node pair in a network by considering the link activeness \cite{MI12B}. They introduced T\_Flow that captures the importance of information flow via active links in social networks. T\_Flow, which uses the same settings as in PropFlow for random walk, considers link weight as well as link activeness when computing transition probabilities.


\section{Future Directions and Challenges}
Although numerous efforts have been made in link prediction, there are still many potential future challenges, and some new open problems require further study. 
Here, we address some possible future research challenges on the link prediction problem.

\begin{itemize}
\item \textbf{Disappearing link prediction} 
Most existing link prediction works focus on links that will appear in the future, only a few works discuss the prediction of links that will disappear in the future \cite{AGJ12}. 
We argue that predicting links that may disappear in the future is also very important. 
This problem has been considered by some researchers. 
However, it is not easy to solve this problem.
Predicting disappearing links is not the inverse problem of predicting appearing or missing links. 
The main reason is that the mechanism of the link dissolution is not same to the mechanism of the link formation.
Therefore, we cannot directly apply the current link prediction methods to predict the disappearing links. 
For example, finding the node pairs with low similarities would not work for this new problem. 
The key of solving this problem should be understand the mechanism of the link dissolution, then reasonable approaches could be designed.   
\end{itemize}

\begin{itemize}
\item \textbf{Link prediction under dynamic nodes} Most current link prediction works have a default assumption: the nodes of social networks are known and will not change in the future. However, in most practical cases, this assumption cannot be satisfied. Social networks are highly dynamic, and a node may join or leave the network. Therefore, link prediction under dynamic nodes is an interesting challenge. Current link prediction methods also cannot work well for this problem.
In social networks such as Twitter and Sina Weibo, many users are never active after a period of time, these users should not be considered in link prediction because they have actually left the social networks. 
A more complicated issue is that there are a lot of fake users controlled by malicious programs, and these users are not real but their social activities are similar to real users, so the link prediction methods should consider the negative influence by fake users.   
\end{itemize}

\begin{itemize}
\item \textbf{Overcoming imbalance} 
Link prediction problem always suffers from extreme imbalance, 
namely, the number of links known to be present is often much less than the number of links known to be absent. 
This imbalance  hampers the effectiveness of many link prediction methods, and it is necessary to  overcome this problem in the future work. 
In addition, a reason of the link prediction being hard lies in the fact of  most interesting linked datasets are very sparse. 
Therefore, it is difficult to build statistical models for link prediction because that the prior probability of a link is typically quite small. 
Then it causes difficulty both in model evaluation and, more importantly, in quantifying the level of confidence in the predictions.
Current link prediction experimental results are usually in very low evaluation metrics values, so the link prediction performances have enough space for improvement. 
\end{itemize}

\begin{itemize}
\item \textbf{Incorporating social theories} 
A large number of methods for link prediction in social networks consider only topological features and attributes, few works take social theory features into consideration. 
Therefore, they are still traditional data mining and learning solutions, and are independent to the social networks. 
The social theories are useful for explaining the mechanisms of social activities. 
Incorporating social theories into the link prediction methods would be promising.
For example, social theories would provide reasonable features for learning-based link prediction methods.
However, some new social theory works could not be noticed by researchers from computer science field.  
Fortunately, some works that incorporating existing social theories into the link prediction methods have shown that the prediction accuracy can be improved. 
More comprehensive works should follow these positive results.
\end{itemize}

\begin{itemize}
\item \textbf{Link prediction in heterogeneous social networks} 
A lot of traditional link prediction methods are limited to homogeneous networks with single-type edges and nodes, but practical social networks usually have multiple relation types and node types. 
For example, a heterogeneous bibliographic network contains nodes such as publications, authors and venues, and edges such as co-author, cite and work-in.
In addition, it needs to predict links across networks in social network applications. 
Some works have drawn attention to this problem.
For example, the probabilistic latent tensor factorization model \cite{GDG11B,GDG12} is proposed to solve this problem.
However, most of these solutions are too complicated and have high time complexities, or are designed for special applications. 
Therefore, it deserves further study for designing more elegant and generalized methods for solving the link prediction problem in heterogeneous social networks.  
\end{itemize}

\begin{itemize}
\item \textbf{Fair evaluation and benchmark datasets} 
Up to now, numerous methods have been proposed to solve the link prediction problem. 
However, the corresponding evaluations are inadequate, which may lead to inappropriate conclusions about the performances of link prediction methods. 
So more fair evaluation for link prediction is very necessary. 
On the other hand, there is no a benchmark dataset for link prediction.
Almost all works need to collect experimental datasets, besides the time-consumed for dataset collection process, other drawback is that these datasets would vary largely in network size and network features.   
Therefore, it is difficult to fairly and systematically compare the performances between new methods and previous ones.
As a result, the performance of a model would be good in some datasets but also would be unsatisfactory in others.
Researchers would select specific datasets to support their works, that would hide the limitation of their link prediction methods. 
In fact, for many problems in computer science, people have built a lot of benchmark datasets, which are very helpful to promote the research development on these problems.
\end{itemize}

\section{Conclusions}

The link prediction is not a new problem in link mining and analysis. 
New link prediction techniques, problems and applications are emerging quickly in recent years. 
This paper attempts to systematically summarize all typical works on the link prediction in social networks.
A category of link prediction techniques and link prediction problems is proposed.
Link prediction techniques are discussed, especially the topology-based metrics and learning-based methods.
Link prediction problems and applications are also presented.
Active research groups are also introduced.
Finally, future directions and challenges are addressed.

\Acknowledgements{This work was supported by 
the National Key Basic Research and Development Program of China (2014CB340702),
the National Natural Science Foundation of China (61170071, 91318301, 61321491, 61472077), 
the China Postdoctoral Science Foundation (2014M560378), 
and the foundation of the State Key Laboratory of Software Engineering(SKLSE).}


\end{document}